\documentclass[prd,twocolumn,showpacs,showkeys,preprintnumbers,floatfix,
  nofootinbib]{revtex4-1}
\usepackage{amsfonts} 
\usepackage{amssymb} 
\usepackage{amsmath} 
\usepackage{subfigure} 
\usepackage{graphicx} 
\usepackage{array} 
\usepackage{dcolumn} 
\usepackage{bm} 
\usepackage{latexsym} 
\usepackage{longtable} 
\usepackage{hyperref} 
\usepackage{color}
\usepackage{mathrsfs}
\usepackage{comment}
\graphicspath{{./figs/}}

\newcommand{\CO}{{\cal O}}

\newcommand{\dmu}{\partial_\mu}

\newcommand{\Tr}{\textrm{Tr}}

\newcommand{\mc}[1]{\mathcal{#1}}
\newcommand{\mr}[1]{\mathrm{#1}}
\renewcommand{\mathcal}[1]{\mathscr{#1}}

\newif\ifContLineOne
\newif\ifContLineTwo
\newif\ifContLineThree

\def\conC#1{\vbox{\ialign{##\crcr
  \ifContLineThree\hrulefill\else\vphantom{\hrulefill}\fi\crcr
  \noalign{\kern3.2pt\nointerlineskip}
  \ifContLineTwo\hrulefill\else\vphantom{\hrulefill}\fi\crcr
  \noalign{\kern3.2pt\nointerlineskip}
  \ifContLineOne\hrulefill\else\vphantom{\hrulefill}\fi\crcr
  \noalign{\nointerlineskip}
  $\hfil\textstyle{\vbox to 14pt{}#1}\hfil$\crcr}}}

\def\DrawLeg#1#2{
  \kern-.2pt              
  \dimen2 =#1             
  \advance\dimen2 by 2pt  
  \dimen3 = 10.6pt        
  \dimen4 =3.6pt          
  \advance\dimen3 by -\dimen2 
  \multiply\dimen4 by #2
  \advance\dimen3 by \dimen4
  \raise\dimen2 \hbox{\vrule height\dimen3 width .4pt} 
  \kern-.2pt}             

\def\begC#1#2{\setbox0 =\hbox{$\textstyle{#2}$}
  \dimen0=.5\wd0 \dimen1=\ht0
  \conC{\hskip\dimen0}
  \count255=#1
  \ifnum\count255 =1 \ContLineOnetrue\else
  \ifnum\count255 =2 \ContLineTwotrue\else
  \ifnum\count255 =3 \ContLineThreetrue\fi\fi\fi
  \DrawLeg{\dimen1}{\count255}
  \conC{\hskip\dimen0}
  \kern-\dimen0\kern-\dimen0 \box0}

\def\endC#1#2{\setbox0 =\hbox{$\textstyle{#2}$}
  \dimen0=.5\wd0 \dimen1=\ht0
  \conC{\hskip\dimen0}
  \count255=#1
  \ifnum\count255 =1 \ContLineOnefalse\else
  \ifnum\count255 =2 \ContLineTwofalse\else
  \ifnum\count255 =3 \ContLineThreefalse\fi\fi\fi
  \DrawLeg{\dimen1}{\count255}
  \conC{\hskip\dimen0}
  \kern-\dimen0\kern-\dimen0 \box0}

\begin{document}

\title{
Taste non-Goldstone, flavor-charged pseudo-Goldstone boson masses in staggered chiral perturbation theory
}
\author{Jon A.~Bailey}
\author{Hyung-Jin Kim}
\author{Weonjong Lee}

\collaboration{SWME Collaboration}

\affiliation{
  Lattice Gauge Theory Research Center, FPRD, and CTP,\\
  Department of Physics and Astronomy,
  Seoul National University, Seoul, 151-747, South Korea
}
\date{\today}
\begin{abstract}
We calculate the masses of taste non-Goldstone pions and kaons in staggered
chiral perturbation theory through next-to-leading order in the standard power
counting.  The results can be used to quantitatively understand taste
violations in existing lattice data generated with staggered fermions and to
extract the $u$, $d$, and $s$ quark masses and Gasser-Leutwyler parameters from
the experimentally observed spectrum.  The expressions for the non-Goldstone
masses contain low-energy couplings unique to the non-Goldstone sector.  With
two exceptions these enter as coefficients of analytic terms; all the new
couplings can be fixed by performing spectrum calculations.  We report one-loop
results for the quenched case and the fully dynamical and partially quenched
1+1+1 and 2+1 flavor cases in the chiral $SU(3)$ and $SU(2)$ theories.  
\end{abstract}
\maketitle
\section{Introduction \label{sec:intr}}

The masses of the up, down, and strange quarks are fundamental
parameters of the Standard Model, and the low-energy couplings (LECs)
of chiral perturbation theory
($\chi$PT)~\cite{Gasser:1983yg,Gasser:1984gg} parametrize the strong
interactions at energies soft compared to the scale of chiral symmetry
breaking, $\Lambda_\mr{\chi}$.  By fitting lattice QCD data to
$\chi$PT, the light quark masses, Gasser-Leutwyler couplings, and
other LECs can be determined~\cite{Colangelo:2010et,Bazavov:2009bb}.

The results of lattice QCD calculations contain discretization effects
that in principle must be taken into account, either before fitting to
$\chi$PT or in the fits themselves.  For sufficiently small lattice
spacings, lattice artifacts perturb the continuum
physics~\cite{Symanzik:1983dc,Symanzik:1983gh}, and the effects of
these perturbations at energies much less than $\Lambda_\mr{\chi}$
can be described by an effective field theory, lattice chiral
perturbation theory~\cite{Lee:1999zxa}.

Staggered fermions possess an exact chiral symmetry at nonzero lattice spacing
and are computationally cheap.  However, in practice discretization effects
known as taste violations are significant even with Symanzik improvement.  In
Ref.~\cite{Lee:1999zxa} Lee and Sharpe introduced the $\chi$PT for a single
flavor of staggered fermion coupled to $SU(3)$ lattice gauge fields.  Working
to leading order (LO) in a dual expansion in the quark masses and lattice
spacing, they showed that the staggered pion spectrum, including taste
violations, respects $SO(4)_T$ taste symmetry and confirmed this prediction of
the $\chi$PT by comparing to lattice data generated by using unimproved and
improved versions of staggered fermions~\cite{Ishizuka:1993mt,Orginos:1998ue}.

Motivated by unsuccessful attempts to describe lattice data by fitting to
standard continuum $\chi$PT~\cite{Bernard:2001av}, Aubin and Bernard
generalized the Lee-Sharpe Lagrangian to multiple flavors to describe the
effects of lattice artifacts, including taste violations, in the
pseudo-Goldstone boson (PGB)
sector~\cite{Bernard:2001yj,Aubin:2003mg,Aubin:2003rg}.  They used the
resulting staggered chiral perturbation theory (S$\chi$PT) to calculate the
masses and decay constants of taste Goldstone pions and kaons (flavor-charged
states) through one loop, including the leading chiral logarithms, which enter
at next-to-leading order (NLO) in the dual
expansion~\cite{Aubin:2003mg,Aubin:2003uc}.  The results were used to
successfully describe lattice data and factored in phenomenologically
successful calculations of quark masses, meson masses, decay constants, form
factors, mixing parameters, and other quantities
~\cite{Bernard:2001yj,Aubin:2004ck,Aubin:2004fs,Aubin:2004ej,Aubin:2005ar,Gray:2005ad,Aubin:2005zv,Okamoto:2005zg,Dalgic:2006dt,Aubin:2006xv,Bernard:2008dn,Bailey:2008wp,Gamiz:2009ku,Bazavov:2009bb,Bae:2010ki,Kim:2011qg}.

In Ref.~\cite{Sharpe:2004is} Sharpe and Van de Water enumerated the
complete NLO Lagrangian of S$\chi$PT and used it to predict
relationships between taste breaking splittings in the PGB masses,
decay constants, and dispersion relations.  The NLO Lagrangian breaks
$SO(4)_T$ to the lattice symmetry group and contributes to the masses
of the PGBs terms at NLO in the dual expansion.

Lattice QCD calculations with staggered fermions are conducted by taking the
fourth root of the fermion determinant to eliminate remnant doubling from the
sea~\cite{Bazavov:2009bb}.  The conjectured relationship of the rooted
staggered theory and QCD has implications that can be numerically tested.  The
rooting is systematically incorporated into S$\chi$PT using the replica
method~\cite{Damgaard:2000gh,Aubin:2003mg,Aubin:2003rg}.  Rooting leads to
unphysical effects at nonzero lattice spacing.  Following the arguments of
Refs.~\cite{Adams:2004wp,Durr:2005ax,Bernard:2006zw,Bernard:2006qt,Bernard:2007yt,Bernard:2007ma,Sharpe:2006re,Creutz:2007rk,Kronfeld:2007ek,Adams:2008db,Creutz:2008nk,Golterman:2008gt,Donald:2011if}, we assume
that the unphysical effects of rooting vanish in the continuum limit and that
S$\chi$PT with the replica method correctly describes the effects of rooting
that enter soft pion processes at nonzero lattice spacing.

Here we calculate the masses of the taste non-Goldstone pions and
kaons through NLO (one loop) in S$\chi$PT.  The results can be used to determine
the up, down, and strange quark masses, the Gasser-Leutwyler
couplings, and other quantities by confronting lattice
data generated with rooted staggered fermions.  Consistency between
lattice data and the S$\chi$PT description of the taste
non-Goldstone sector would constitute additional numerical evidence
for the conjectured relationship between the rooted staggered theory
and QCD.

In Sec.~\ref{sec:review} we review the formalism of S$\chi$PT, in
Sec.~\ref{sec:se} we calculate the self-energies of the taste non-Goldstone
states, Sec.~\ref{sec:res} contains the resulting one-loop corrections to the
masses, and in Sec.~\ref{sec:disc} we discuss the results and note directions
for future work.  Appendices~\ref{app:pcf}, \ref{app:pattern}, \ref{app:eg},
and \ref{sum_theta} respectively contain a derivation of the power counting through NLO, a discussion of the taste symmetry breaking induced by NLO analytic terms, details of the loop calculations, and details of the calculation of the
coefficients of the chiral logarithms.


\section{\label{sec:review}Staggered chiral perturbation theory}
Here we briefly review the formulation of S$\chi$PT~\cite{Lee:1999zxa,Aubin:2003mg}, recalling relevant differences between the staggered theory and its continuum counterpart~\cite{Gasser:1983yg,Gasser:1984gg}.  
For simplicity we consider the 4+4+4 theory of Refs.~\cite{Aubin:2003rg,Aubin:2003mg}.
The symmetries and degrees of freedom of S$\chi$PT are recalled in Sec.~\ref{dofs}, the extension of the power counting in Sec.~\ref{sec:pow}, and the Lagrangian in Sec.~\ref{lag}.

Pedagogical treatments of lattice $\chi$PT are given in
Refs.~\cite{Golterman:2009kw,Sharpe:2006pu}.  Investigations of the
foundations of S$\chi$PT were reported in
Refs.~\cite{Bernard:2006zw,Bernard:2006qt,Bernard:2007yt,Bernard:2007ma}.

\subsection{\label{dofs} Group theory and degrees of freedom}

In the continuum limit, the chiral symmetry of the 4+4+4 theory is
$SU(12)_L\times SU(12)_R$.  Assuming spontaneous breaking to the
vector subgroup, the 143 pseudo-Goldstone bosons (PGBs) in the adjoint
irrep of $SU(12)_V$ can be classified according to the continuum
flavor-taste subgroup $SU(3)_F\times SU(4)_T$:
\begin{eqnarray}
SU(12)_V&\supset & SU(3)_F\times SU(4)_T\\
\mathbf{143}&\rightarrow &\mathbf{(8,\ 15)\oplus(8,\ 1)\oplus(1,\ 15)}\label{su12tosu3xsu4}
\end{eqnarray}
Discretization effects break the direct product of the continuum
chiral symmetry and Euclidean rotations to a direct product of the
lattice chiral symmetry and hypercubic rotations~\cite{Aubin:2003mg}:
\begin{eqnarray}
& U(1)_V \times SU(12)_L\times SU(12)_R\times SO(4)&\\
&\xrightarrow{a\neq0} U(3)_l\times U(3)_r 
\times(\Gamma_4\rtimes SW_{4,\mathrm{diag}})&
\end{eqnarray}
$U(3)_l\times U(3)_r$ is the lattice chiral symmetry of three flavors
of staggered fermions.  Its appearance ensures that a nonet of the
PGBs becomes massless in the chiral limit even at nonzero lattice
spacing; by definition, these are the taste Goldstone states.  The
$U(1)_a$ is not to be confused with the anomalous axial symmetry of
the theory; the flavor-taste singlet meson receives a large
contribution to its mass from the anomaly and is not among the PGBs.

The Clifford group $\Gamma_4$ is a subgroup of taste $SU(4)_T$;
$\Gamma_4$ is generated by the Hermitian, $4\times 4$ matrices
$\xi_\mu$, $\{\xi_\mu,\xi_\nu\}=2\delta_{\mu\nu}$~\cite{Lee:1999zxa}.
$SW_{4,\mathrm{diag}}$ is the group of hypercubic rotations embedded
in the diagonal of the direct product of Euclidean rotations and the
remnant taste $SO(4)_T$ that emerges at energies soft compared to the
scale of chiral symmetry breaking~\cite{Lee:1999zxa}:
\begin{eqnarray}
SO(4)\times SU(4)_T&\xrightarrow{a\ne0}& SW_{4,\mathrm{diag}}\\
& \underset{\phantom{\chi}p\ll\Lambda_\chi}{\subset} & SO(4) \times SO(4)_T
\label{eq:schpt-lo}
\end{eqnarray}
where Eq.~(\ref{eq:schpt-lo}) represents the symmetry of the staggered
chiral Lagrangian at leading order.

The emergence of $SO(4)_T$ implies degeneracies among the PGBs.  The
fundamental rep of $SU(4)_T$ is a spinor under $SO(4)_T$, and the
$SU(4)_T$ adjoint and singlet of Eq.~(\ref{su12tosu3xsu4}) fall into
five irreps of $SO(4)_T$~\cite{Aubin:2003mg}:
\begin{eqnarray}
SU(4)_T&\supset & SO(4)_T\\
\mathbf{15}&\rightarrow & P\oplus A\oplus T\oplus V\\
\mathbf{1}&\rightarrow & I
\end{eqnarray}
The $SO(4)_T$ irreps are the pseudoscalar, axial vector, tensor,
vector, and singlet (or scalar), respectively.  The flavor-nonet
taste-pseudoscalar PGBs are the taste Goldstone states.  Among them
are the pions and kaons of Refs.~\cite{Aubin:2003rg,Aubin:2003mg}.
The taste singlet $\eta^\prime$ receives a large contribution to its
mass from the anomaly and can be integrated out of the theory.

At nonzero quark masses, the continuum chiral symmetry is softly broken to
$SU(12)_V$, $SU(8)_V\times SU(4)_V$, or $SU(4)_V\times SU(4)_V\times SU(4)_V$
for three degenerate, two degenerate, or three non-degenerate flavors,
respectively.  Noting the anomaly contribution in the taste singlet sector and
assuming the taste singlet PGBs are degenerate with their physical counterparts
in the continuum limit, we can use the isospin, strangeness, and
continuum vector symmetries to deduce the degeneracies between the remaining
(taste non-singlet PGBs) and the physical states.  

In doing so we consider the target continuum theory with 1+1+1 flavors, in
which there are 12 valence quarks (and 12 ghost quarks), but the fourth root
has reduced the number of sea quarks from 12 flavors to three.  We also
restrict our attention to PGBs constructed exclusively of valence quarks.  The
resulting deductions from symmetry represent one of the simplest testable
implications of the correctness of the rooting conjecture.  We can also use
them to check our S$\chi$PT calculation of the masses of the taste
non-Goldstone pions and kaons.  
%
%
%
%

\subsection{\label{sec:pow}Power counting}

The standard power counting of S$\chi$PT~\cite{Lee:1999zxa} is a
straightforward generalization of that in the continuum
theory~\cite{Weinberg:1978kz,Gasser:1984gg}.  The Lagrangian is
expanded in a series of local interactions perturbing the low-energy
theory about the chiral and continuum limits, and observables are
calculated in a dual expansion in the quark masses and lattice
spacing.

The order of an operator in the Lagrangian corresponds to the number of derivatives, quark mass factors, and lattice spacing factors in the operator.  The symmetries of staggered fermions ensure the leading lattice artifacts are $\CO(a^2)$, and derivatives always appear in pairs~\cite{Lee:1999zxa}.  Let $n_{p^2}$, $n_{m}$, and $n_{a^2}$ be the number of derivative pairs, quark mass factors, and (squared) lattice spacing factors in an interaction.  The general form of the Lagrangian is
\begin{eqnarray}
\mathcal{L}&=&\sum_{n=1}^\infty\mathcal{L}_{2n}=\mathcal{L}_2 + \mathcal{L}_4 + \dots\label{LagExp}\\
&=&\sum_{n=1}^\infty\mathcal{L}_{\mathrm{N}^{n-1}\mathrm{LO}}=\mathcal{L}_\mathrm{LO}+\mathcal{L}_\mathrm{NLO}+\dots,
\end{eqnarray}
where $n=n_{p^2}+n_{m}+n_{a^2}$ and $\mathcal{L}_{\mathrm{N}^{n-1}\mathrm{LO}}\equiv\mathcal{L}_{2n}$.

This organization of the Lagrangian is consistent with the expectation
that contributions at leading non-trivial order will be
\begin{equation}
\CO(p^2/\Lambda_\chi^2)
\approx \CO(m_q/\Lambda_\chi)
\approx \CO(a^2\Lambda_\chi^2).\label{eq:count}
\end{equation}
This power counting is appropriate to data generated
on the MILC asqtad coarse lattices ($a \approx 0.12$ fm); on finer lattices or with more improved versions of the staggered action, the discretization effects are often smaller.

Feynman graphs are functions of external momenta $p_i$, the quark
masses $m_q$, and the lattice spacing $a^2$:
\begin{equation}
\mathcal{M}(p_i,m_q,a^2),
\end{equation}
where the amplitude $\mathcal{M}$ is related to the $\mathbb{S}$-matrix
as follows:
\begin{equation}
\mathbb{S} \sim \delta^4(\sum_i p_i) \mathcal{M}.
\end{equation}
Rescaling $p_i$, $m_q$, and $a^2$ to smaller values in accord with the power counting in Eq.~(\ref{eq:count}), we have
\begin{equation}
\mathcal{M}(p_i,m_q,a^2)\rightarrow\mathcal{M}(\sqrt{\varepsilon}p_i,\varepsilon m_q,\varepsilon a^2),
\end{equation}
which leads to~\cite{Scherer:2002tk}
\begin{eqnarray}
&\mathcal{M}(\sqrt{\varepsilon}p_i,\varepsilon m_q,\varepsilon a^2)
=\varepsilon^{D}\mathcal{M}(p_i,m_q,a^2),&\label{eq:scale}\\
&D=1 + {\displaystyle \sum_{n=1}^\infty(n-1)N_{2n}} + N_L.&\label{eq:power}
\end{eqnarray}
A derivation of Eq.~\eqref{eq:power} is given in Appendix \ref{app:pcf}.
$N_L$ is the number of loops in the graph, and $N_{2n}$ is the number
of vertices from operators in $\mathcal{L}_{2n}$.  From
Eqs.~(\ref{eq:scale}) and (\ref{eq:power}), we see that loops and
diagrams with vertices from higher order interactions are suppressed
at small momenta, quark masses, and lattice spacings.

For any given observable we first consider all graphs with $D=1$ (LO), then those with $D=2$ (NLO), and so on.  At leading order ($D=1$) the only solutions to Eq.~(\ref{eq:power}) have $N_{2n}=0$ for $n\geq2$ and $N_L=0$; {\it i.e.}, only tree graphs with vertices from the LO Lagrangian are allowed.  At NLO ($D=2$), the solutions have $N_{2n}=0$ for $n\geq3$ and either $N_4=1$ or $N_L=1$; one-loop graphs with vertices from the LO Lagrangian and tree graphs with at most one vertex from the NLO Lagrangian are allowed.  In Sec.~\ref{sec:se} we use these observations to write down the graphs contributing to the PGB self-energies through NLO in the dual expansion.

\subsection{\label{lag}Lagrangian}

The Lagrangian is constructed of the PGB fields $\phi$, quark mass matrix $M$, derivatives, and taste matrices $\xi_\mu$ in accord with the symmetries of the terms in the effective continuum Symanzik action~\cite{Lee:1999zxa,Aubin:2003mg,Sharpe:2004is}.  The exponential parametrization is a convenient way to include the PGBs.
\begin{equation}
\Sigma=e^{i\phi/f},\quad SU(12)_L\times SU(12)_R:\ \Sigma\rightarrow L\Sigma R^\dagger 
\end{equation}
where $L,\ R\in SU(12)_{L,R}$ and 
\begin{eqnarray}
\phi&=&\sum_a \phi^a\otimes T^a\\
\phi^a&=&{
\begin{pmatrix}
U_a & \pi^+_a & K^+_a \\
\pi^-_a & D_a & K^0_a \\
K^-_a & \bar K^0_a & S_a
\end{pmatrix}}\\
T^a&\in&\{\xi_5,\ i\xi_{\mu5},\ i\xi_{\mu\nu}(\mu<\nu),\ \xi_\mu, \xi_I\}.
\label{eq:T^a}
\end{eqnarray}
The index $a$ runs over the 16 PGB tastes in the $\mathbf{15}$ and $\mathbf{1}$ of $SU(4)_T$, the $\phi^a$ are Hermitian $3\times 3$ matrices, and the $T^a$ are Hermitian $4\times 4$ generators of $U(4)_T$, chosen (up to phases) as members of the Clifford algebra generated by the matrices $\xi_\mu$.  With this choice for the $T^a$, the $SO(4)_T$ quantum numbers of the PGBs are explicit.

We follow Refs.~\cite{Aubin:2003mg,Sharpe:2004is} in including the flavor-taste $SU(12)_V$ singlet in the Lagrangian, so $\Sigma\in U(12)$.  An additional mass term in the Lagrangian accounts for the anomaly contribution to the mass of the $SU(12)_V$ singlet.
Taking this mass correction to infinity at the end of the calculation properly decouples the $SU(12)_V$ singlet and yields the desired results~\cite{Sharpe:2000bc,Sharpe:2001fh}.

At leading order in the expansion of the Lagrangian, there are three classes of interactions:  operators with $(n_{p^2},n_m,n_{a^2})=(1,0,0)$, $(0,1,0)$, and $(0,0,1)$.  We have
\begin{eqnarray}
\mathcal{L}_\mathrm{LO} &=&\frac{f^2}{8} \Tr(\partial_{\mu}\Sigma \partial_{\mu}\Sigma^{\dagger}) - 
\frac{1}{4}\mu f^2 \Tr(M\Sigma+M\Sigma^{\dagger}) \nonumber\\
&+& \frac{2m_0^2}{3}(U_I + D_I + S_I)^2 + a^2 (\mathcal{U+U^\prime})
\label{F3LSLag}
\end{eqnarray}
where
\begin{equation}
M=
\begin{pmatrix}
m_u & 0 & 0 \\
0   & m_d & 0 \\
0 & 0 & m_s 
\end{pmatrix}
\otimes \xi_I,
\end{equation}
$\xi_I$ is the identity matrix in taste space, and the trace (in flavor-taste space) is ordinary; we use the replica method of Damgaard and Splittorff to generalize the results of the 4+4+4 theory to the partially quenched case~\cite{Damgaard:2000gh}.

The term proportional to $m_0^2$ is the contribution from the anomaly, and the potentials $\mathcal{U}$ and $\mathcal{U^\prime}$ break $SU(4)_T$ to $SO(4)_T$.  They are
\begin{eqnarray}
 -\mathcal{U}  = & \mbox{ } & C_1\Tr(\xi^{(n)}_5\Sigma\xi^{(n)}_5\Sigma^{\dagger}) \nonumber \\
 &+&C_6\ \sum_{\mu<\nu} \Tr(\xi^{(n)}_{\mu\nu}\Sigma \xi^{(n)}_{\nu\mu}\Sigma^{\dagger}) \nonumber \\
 &+&C_3\tfrac{1}{2} \sum_{\nu}[ \Tr(\xi^{(n)}_{\nu}\Sigma \xi^{(n)}_{\nu}\Sigma) + h.c.] \nonumber \\
 &+&C_4\tfrac{1}{2} \sum_{\nu}[ \Tr(\xi^{(n)}_{\nu 5}\Sigma \xi^{(n)}_{5\nu}\Sigma) + h.c.]
\label{TBPot1}
\end{eqnarray}
\begin{eqnarray}
 -\mathcal{U}^\prime=&\mbox{ }&C_{2V}\tfrac{1}{4}\sum_{\nu}[\Tr(\xi^{(n)}_{\nu}\Sigma)\Tr(\xi^{(n)}_{\nu}\Sigma)  + h.c.] \nonumber \\
 &+& C_{2A}\tfrac{1}{4} \sum_{\nu}[ \Tr(\xi^{(n)}_{\nu5}\Sigma)\Tr(\xi^{(n)}_{5\nu}\Sigma)  + h.c.] \nonumber \\*
 &+& C_{5V}\tfrac{1}{2} \sum_{\nu}[ \Tr(\xi^{(n)}_{\nu}\Sigma)
 \Tr(\xi^{(n)}_{\nu}\Sigma^{\dagger})]\nonumber \\
 &+& C_{5A}\tfrac{1}{2} \sum_{\nu}[ \Tr(\xi^{(n)}_{\nu5}\Sigma)
 \Tr(\xi^{(n)}_{5\nu}\Sigma^{\dagger}) ]
\label{TBPot2}
\end{eqnarray}
where $T^{a(n)}=T^{a(3)}\equiv I_3\otimes T^a$ in the 4+4+4 theory and $T^a$ is
given in Eq.~\ref{eq:T^a}. $I_3$ is the identity matrix in flavor space.  The
potentials are derived by mapping the operators of the mass dimension six
effective continuum Symanzik action into the operators of $\chi$PT.
The remnant taste symmetry of Lee and Sharpe emerges because contributions to
the potential from $SO(4)_T$-breaking operators in the Symanzik action are
suppressed in the low-energy effective field theory by powers of the
four-momenta of the PGBs.  The derivation of the potentials and the restoration
of taste $SO(4)_T$ symmetry are described in detail in
Refs.~\cite{Lee:1999zxa,Aubin:2003mg,Sharpe:2004is}.

At NLO, the Lagrangian operators fall into six classes:  $(n_{p^2},n_m,n_{a^2})=(2,0,0)$, $(0,2,0)$, $(1,1,0)$, $(1,0,1)$, $(0,1,1)$, and $(0,0,2)$.  The first three contain terms analogous to those in the Gasser-Leutwyler Lagrangian~\cite{Gasser:1984gg}.  The last three contain the terms enumerated by Sharpe and Van de Water~\cite{Sharpe:2004is}.  The Gasser-Leutwyler terms of S$\chi$PT that contribute to the PGB masses at NLO are
\begin{align}
\mathcal{L}_\mathrm{GL}=&\phantom{+}L_4\Tr(\dmu\Sigma^\dagger\dmu\Sigma)\Tr(\chi^\dagger\Sigma+\chi\Sigma^\dagger)\nonumber\\
&+L_5\Tr(\dmu\Sigma^\dagger\dmu\Sigma(\chi^\dagger\Sigma+\Sigma^\dagger\chi))\nonumber\\
&-L_6[\Tr(\chi^\dagger\Sigma+\chi\Sigma^\dagger)]^2\nonumber\\
&-L_8\Tr(\chi^\dagger\Sigma\chi^\dagger\Sigma) + h.c.
\label{eq:GL}
\end{align}
where $\chi=2\mu M$.

Many operators in the Sharpe-Van de Water Lagrangian contribute at NLO, but only a handful break the remnant taste $SO(4)_T$ to the hypercubic subgroup $SW_4$ of the lattice theory~\cite{Sharpe:2004is}.
We use the symmetries of the Sharpe-Van de Water terms to deduce the form of their contributions to the masses; as discussed in Appendix~\ref{app:pattern}, the explicit results of Sharpe and Van de Water for the $SO(4)_T$-breaking contributions to the flavor-charged PGB dispersion relations restrict the number of independent parameters in these contributions to only three.

\section{\label{sec:se}Self-energies of flavor-charged pseudo-Goldstone bosons}
%
%
The symmetries protect the flavor-charged PGBs from mixing.  For reasons
discussed in Sec.~\ref{sec:diag} below, here we describe the calculation in the rest
frame.  In terms
of the self-energy $\Sigma(p_4^2)$ at $\vec{p}=0$ of the state
$\phi_{xy}^a\ (x\neq y,\ x,y\in\{u,d,s\})$,
\begin{equation}
M_\phi^2 = m^2_\phi + \Sigma(-M^2_\phi),\label{eq:defmass}
\end{equation}
where $m_\phi$ is the tree-level (LO) mass, and $M_\phi$ is the
(exact) mass to all orders.  Noting that the perturbative expansion of
$\Sigma(p_4^2)$ begins at NLO and expanding $\Sigma(p_4^2)$ in a Taylor
series around $p_4^2=-m^2_\phi$ gives
\begin{eqnarray*}
M_\phi^2&=&m^2_\phi + \Sigma(-m^2_\phi) 
- \Sigma(-M^2_\phi)\Sigma^\prime(-m^2_\phi) + \dots\\ 
&=&m^2_\phi + \Sigma(-m^2_\phi) + \mathrm{NNLO},
\end{eqnarray*}
and the NLO correction to the mass is the leading contribution to $\Sigma(-m^2_{\phi})$.

In Sec.~\ref{sec:diag} we consider the Feynman graphs entering the expansion of
the self-energies at NLO.  In Sec.~\ref{sec:calc} we outline the calculation of
these graphs, and in Sec.~\ref{sec:res444}, we present a condensed version of
the results for the 4+4+4 theory.

\subsection{\label{sec:diag}Diagrammatic expansion}

\begin{figure}
\begin{center}
\includegraphics[width=5cm]{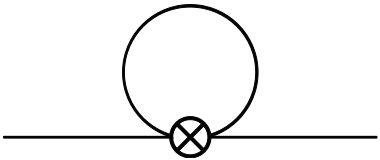}
\caption{At next-to-leading order, tadpole graphs contribute the leading chiral logarithms.  The vertices are from the leading order Lagrangian of Eq.~(\ref{F3LSLag}), and the propagator represents the connected part (first term) of Eq.~(\ref{prop}).}
\label{fig:scon}
\end{center}
\end{figure}

\begin{figure}
\begin{center}
\includegraphics[width=5cm]{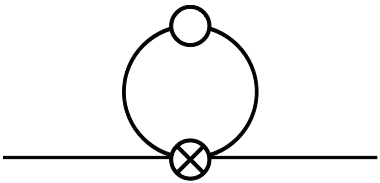}
\caption{Disconnected tadpoles enter in the flavor-neutral, taste-singlet, -vector, and -axial channels.  The open circle represents the second term of Eq.~(\ref{prop}).}
\label{fig:sdisc}
\end{center}
\end{figure}

\begin{figure}
\begin{center}
\includegraphics[width=5cm]{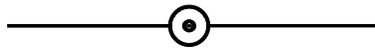}
\caption{At next-to-leading order, tree-level graphs contribute terms analytic in the quark masses and (squared) lattice spacing.  The vertices are from the Gasser-Leutwyler and Sharpe-Van de Water Lagrangians of Eq.~(\ref{eq:GL}) and Ref.~\cite{Sharpe:2004is}.}
\label{fig:sanal}
\end{center}
\end{figure}

The power counting of Sec.~\ref{sec:pow}, the Lagrangian of Sec.~\ref{lag}, and
the definition of $\Sigma$ constrain the diagrams entering the NLO mass
corrections to three types.  Expanding the LO Lagrangian through $\CO(\phi^4)$
and the NLO Lagrangian through $\CO(\phi^2)$, we write
\begin{equation}
\Sigma(p_4^2) = \frac{1}{(4\pi f)^2}\left[ \sigma^{con}(p_4^2)
 + \sigma^{disc}(p_4^2) \right] + \sigma^{anal}(p_4^2) + \dots\label{SelfEDiv}
\end{equation}
where $\sigma^{con}$ corresponds to the sum of connected tadpole diagrams
(Fig.~\ref{fig:scon}), $\sigma^{disc}$ corresponds to the sum of disconnected
tadpoles (Fig.~\ref{fig:sdisc}), and $\sigma^{anal}$ corresponds to the sum of
tree-level diagrams (Fig.~\ref{fig:sanal}).  The tree-level diagrams are
analytic in the quark masses and (squared) lattice spacing, while the loops
contribute the leading chiral logarithms.

The 4-point vertices in the tadpole graphs are from the $\CO(\phi^4)$ terms in
the LO Lagrangian of Eq.~(\ref{F3LSLag}), and the 2-point vertices in the
tree-level diagrams are from the $\CO(\phi^2)$ terms in the NLO Lagrangian of
Eq.~(\ref{eq:GL}) and the NLO Lagrangian of Sharpe and Van de
Water~\cite{Sharpe:2004is}.  The disconnected propagators (in the graphs of
Fig.~\ref{fig:sdisc}) are from quark-level disconnected contributions to the
tree-level, flavor-neutral propagators in the taste singlet, axial, and vector
channels~\cite{Aubin:2003mg}.

The one-loop graphs break taste $SU(4)_T$ to the remnant taste $SO(4)_T$ of
Ref.~\cite{Lee:1999zxa}, the tree-level graphs from the Gasser-Leutwyler
Lagrangian respect $SU(4)_T$, and the tree-level graphs from the Sharpe-Van de
Water Lagrangian break $SU(4)_T$ in two stages:  Terms of $\CO(a^2m_q)$ and
$\CO(a^4)$ break $SU(4)_T$ to $SO(4)_T$, while terms of $\CO(a^2p^2)$ break the
spacetime-taste symmetry $SO(4)\times SO(4)_T$ to $SW_\mathrm{4,diag}$~\cite{Sharpe:2004is}.  

The one-loop graphs respect spacetime $SO(4)$ rotations, and the corresponding
contributions to the self-energies, $\sigma^{con}$ and $\sigma^{disc}$, are
functions of $p^2$.  The $SO(4)_T$-breaking analytic terms of $\CO(a^2p^2)$,
however, cannot in general be written as functions of $p^2$:  The dispersion
relations are distorted at nonzero lattice spacing by the taste violations.  To
extract the masses one may consider the self-energies in the rest frame.  In
this case the self-energy may be written as a function of $p_4^2$, the square
of the temporal component of the momentum.  In Appendix~\ref{app:pattern} we
recall the form of the $SO(4)_T$-breaking corrections to the dispersion
relations~\cite{Sharpe:2004is}.

\subsection{\label{sec:calc}Calculation in 4+4+4 theory}

For the 4+4+4 theory of Ref.~\cite{Aubin:2003mg}, we outline the calculation of the graphs in Figs.~\ref{fig:scon}, \ref{fig:sdisc}, and \ref{fig:sanal}.  After writing down the propagators and vertices (Sec.~\ref{sec:pav}), we present results for the loops corresponding to each class of vertices (Sec.~\ref{sec:vres}).  These intermediate results are readily checked against the taste Goldstone case~\cite{Aubin:2003mg}.  Sec.~\ref{sec:res444} contains a condensed version of these results, which can be straightforwardly generalized to the partially quenched 1+1+1 theory and other cases of interest (Sec.~\ref{sec:res}).

We calculate the loops without extracting the vertices by summing over the
flavor and taste indices in the $\CO(\phi^4)$ terms in the Lagrangian.  Instead
we combine the expressions for the (tree-level) propagators of the flavor-charged and flavor-neutral PGBs and perform the Wick contractions
before summing over the $\CO(\phi^4)$ vertices.  In Appendix~\ref{app:eg} we
detail the calculation of the contributions from the mass and $a^2\mc{U}$
vertices.

\subsubsection{\label{sec:pav}Propagators and vertex classes}
Expanding the LO Lagrangian of Eq.~\eqref{F3LSLag} through $\CO(\phi^2)$ yields the propagators~\cite{Sharpe:2000bc,Aubin:2003mg}.  They are
\begin{equation}
\langle\phi^a_{ij}\phi^b_{kl}\rangle=\delta^{ab}\left(\delta_{il}\delta_{jk}\frac{1}{q^2+\frac{1}{2}(I_a+J_a)}+\delta_{ij}\delta_{kl}D^a_{il}\right)\label{prop}
\end{equation}
where $i,j,k,l\in\{u,d,s\}$ are flavor $SU(3)_F$ indices, $a,b$ are taste indices in the adjoint irrep, and 
\begin{align}
D^a_{il}&=-\frac{\delta_a}{(q^2+I_a)(q^2+L_a)}\nonumber \\
&\times\frac{(q^2+U_a)(q^2+D_a)(q^2+S_a)}{(q^2+\pi^0_a)(q^2+\eta_a)(q^2+\eta^\prime_a)},\label{D_piece}
\end{align}
where
\begin{align}
\delta_I = 4m_0^2/3,\quad&\delta_{\mu\nu} = 0,\quad\delta_5 = 0 \\
\delta_\mu = a^2\delta_V^\prime,&\quad\delta_{\mu5} = a^2\delta_A^\prime,\label{hpV_hpA}
\end{align}
and the names of mesons denote the squares of their tree-level masses.
For $X\in\{I,J,L,U,D,S\}$,
\begin{align}
X_a\equiv m_{X_a}^2 = 2\mu m_x + a^2\Delta_a,\label{eq:fntm}
\end{align}
where $m_x$ is the mass of the quark of flavor $x\in\{i,j,l,u,d,s\}$, while for
$X\in\{\pi^0,\eta,\eta^\prime\}$, the squares of the tree-level meson masses are the eigenvalues
of the matrix
\begin{equation}\label{444_mass_matrix}
	\begin{pmatrix}
	U_a +\delta_a & \delta_a & \delta_a \\
	\delta_a & D_a +\delta_a & \delta_a \\
	\delta_a & \delta_a & S_a +\delta_a
	\end{pmatrix}.
\end{equation}
The squared tree-level mass of a flavor-charged meson $\phi^t_{xy}\ (x\ne y)$ is
\begin{align}
P_t\equiv\frac{1}{2}(X_t+Y_t)=\mu(m_x+m_y)+a^2\Delta_t,\label{eq:tm}
\end{align}
where $X\ne Y\in\{U,D,S\}$ and $x\ne y\in\{u,d,s\}$.  The hairpin couplings
$\delta_{V,A}^\prime$ and taste splittings $\Delta_a$ are combinations of the
couplings of the LO Lagrangian~\cite{Aubin:2003mg}:
%
%
\begin{align}
\delta_V^\prime &= \frac{16}{f^2}(C_{2V}-C_{5V})\label{hpdef}\\
\delta_A^\prime &= \frac{16}{f^2}(C_{2A}-C_{5A})\label{hpdefA}
\end{align}
and
\begin{align}
\Delta_I &= \frac{16}{f^2}(4C_3+4C_4)\label{Itsplitdef}\\
\Delta_V &= \frac{16}{f^2}(C_1+3C_6+C_3+3C_4)
\end{align}
\begin{align}
\Delta_T &= \frac{16}{f^2}(4C_6+2C_3+2C_4)\\
\Delta_A &= \frac{16}{f^2}(C_1+3C_6+3C_3+C_4)\\
\Delta_P &= 0\label{Ptsplitdef}.
\end{align}
We note the symmetry of Eqs.~\eqref{hpdef}-\eqref{Ptsplitdef} under
simultaneous interchange of vector and axial taste labels
$(V\leftrightarrow A)$ and the coefficients $C_3$ and $C_4$.  The
axial taste matrices $i\xi_{\mu5}$ generate the Clifford algebra; as a
consequence, the LO Lagrangian is invariant under simultaneous
interchange of the fields $\phi^\mu\leftrightarrow -\phi^{\mu5}$ and
the coefficients in the pairs $(C_3,C_4)$, $(C_{2V},C_{2A})$, and
$(C_{5V},C_{5A})$.  

The minus sign arises in the unitary
transformation connecting the bases $\{\xi_\mu\}$ and
$\{i\xi_{\mu5}\}$:  For $U$ such that
\begin{align}
U\xi_\mu U^\dagger &= i\xi_{\mu5}\,,\\
Ui\xi_{\mu5}U^\dagger &=-\xi_\mu\,.\label{eq:minus}
\end{align}
$U$ is unique up to a phase:
\begin{align}
U=e^{i\theta}e^{-i\frac{\pi}{4}\xi_5}
=\frac{1}{\sqrt{2}}e^{i\theta}(\xi_I-i\xi_5),\label{eq:transf}
\end{align}
and the other taste generators are invariant under $U$.  

Noting the diagrammatic expansion and the conservation of $SO(4)_T$, we see
that taste vector and axial fields must always appear in pairs in the
calculation of the self-energies through NLO, and the minus sign in
Eq.~\eqref{eq:minus} is inconsequential.  The results in the taste singlet,
tensor, and Goldstone (pseudoscalar) channels must be invariant under
interchange of the coefficients in the above pairs, while the results in the
taste vector and axial channels must switch.  This symmetry provides a check 
at each stage of the calculation.
%
%
%
%
\begin{widetext}
Expanding the LO Lagrangian of Eqs.~(\ref{F3LSLag}), (\ref{TBPot1}), and (\ref{TBPot2}) and keeping terms of $\CO(\phi^4)$ gives 11 classes of vertices.  From the kinetic energy we have two classes, from the mass terms we have one, and from the potential $a^2\mathcal{U}$, we have four:
\begin{eqnarray}
\frac{f^2}{8} \Tr(\dmu\Sigma \dmu\Sigma^{\dagger})&=&\phantom{-}\frac{1}{48f^2}\,\tau_{abcd}\,(\dmu\phi^a_{ij}\,\phi^b_{jk}\,\partial_\mu\phi^c_{kl}\,\phi^d_{li}
-\dmu\phi^a_{ij}\,\dmu\phi^b_{jk}\,\phi^c_{kl}\phi^d_{li})+ \dots \label{KE4pt} \\
-\frac{1}{4}\mu f^2 \Tr(M\Sigma+M\Sigma^{\dagger})&=&-\frac{\mu}{48f^2}\,\tau_{abcd}\,m_i\,\phi^a_{ij}\phi^b_{jk}\phi^c_{kl}\phi^d_{li} + \dots \label{mass4pt}\\
-a^2C_1\Tr(\xi^{(n)}_5\Sigma\xi^{(n)}_5\Sigma^{\dagger})&=&-\frac{a^2C_1}{12f^4}\,(\tau_{abcd}+3\tau_{5ab5cd}-4\tau_{5a5bcd})\,\phi^a_{ij}\phi^b_{jk}\phi^c_{kl}\phi^d_{li} + \dots \label{C1_4pt}\\
-a^2C_6
\sum_{\mu<\nu} \Tr(\xi^{(n)}_{\mu\nu}\Sigma \xi^{(n)}_{\nu\mu}\Sigma^{\dagger})&=&
-\frac{a^2C_6}{12f^4}\,\sum_{\mu<\nu}(\tau_{abcd}+3\tau_{\mu\nu,ab,\mu\nu,cd}-4\tau_{\mu\nu,a,\mu\nu,bcd})\,\phi^a_{ij}\phi^b_{jk}\phi^c_{kl}\phi^d_{li} + \dots \label{C6_4pt}\\
-a^2C_3
\frac{1}{2} \sum_{\nu}[ \Tr(\xi^{(n)}_{\nu}\Sigma \xi^{(n)}_{\nu}\Sigma) + h.c.]&=&
-\frac{a^2C_3}{12f^4}\,\sum_\nu(\tau_{abcd}+3\tau_{\nu ab\nu cd}+4\tau_{\nu a\nu bcd})\,\phi^a_{ij}\phi^b_{jk}\phi^c_{kl}\phi^d_{li} + \dots \label{C3_4pt}\\
-a^2C_4
\frac{1}{2} \sum_{\nu}[ \textrm{Tr}(\xi^{(n)}_{\nu 5}\Sigma \xi^{(n)}_{5\nu}\Sigma) + h.c.]&=&
-\frac{a^2C_4}{12f^4}\,\sum_\nu(\tau_{abcd}+3\tau_{\nu5,ab,\nu5,cd}+4\tau_{\nu5,a,\nu5,bcd})\,\phi^a_{ij}\phi^b_{jk}\phi^c_{kl}\phi^d_{li} + \dots\label{C4_4pt}
\end{eqnarray}
where the indices $a,b,c,d$ run over the 16 tastes in the $\mathbf{15}$ and $\mathbf{1}$ of $SU(4)_T$, and
$\tau_{abc\cdots}\equiv\Tr(T^aT^bT^c\cdots)$
are traces of products of (Hermitian) taste matrices.
The four operators of the potential $a^2\mathcal{U}^\prime$ each give one class:
\begin{align}
-a^2\sum_\nu\Bigl(C_{2V} & \tfrac{1}{4}[\Tr(\xi^{(n)}_{\nu}\Sigma)\Tr(\xi^{(n)}_{\nu}\Sigma) + h.c.]
-C_{5V}\tfrac{1}{2} [ \Tr(\xi^{(n)}_{\nu}\Sigma) \Tr(\xi^{(n)}_{\nu}\Sigma^{\dagger})]\Bigr)\nonumber \\
&=-\frac{2a^2}{3f^4}(C_{2V}-C_{5V})\,\sum_{\nu}\tau_{\nu abc}\,\phi^\nu_{ii}\,\phi^a_{jk}\phi^b_{kl}\phi^c_{lj} + \dots \label{delV_4pt}\\
-a^2\sum_\nu\Bigl(C_{2A} & \tfrac{1}{4} [ \Tr(\xi^{(n)}_{\nu5}\Sigma)\Tr(\xi^{(n)}_{5\nu}\Sigma) + h.c.]
-C_{5A}\tfrac{1}{2} [ \Tr(\xi^{(n)}_{\nu5}\Sigma) \Tr(\xi^{(n)}_{5\nu}\Sigma^{\dagger}) ]\Bigr)\nonumber \\
&=-\frac{2a^2}{3f^4}(C_{2A}-C_{5A})\,\sum_{\nu}\tau_{\nu5, abc}\,\phi^{\nu5}_{ii}\,\phi^a_{jk}\phi^b_{kl}\phi^c_{lj} + \dots \label{delA_4pt}\\
-a^2\sum_\nu\Bigl(C_{2V}&\tfrac{1}{4}[\Tr(\xi^{(n)}_{\nu}\Sigma)\Tr(\xi^{(n)}_{\nu}\Sigma) + h.c.]
+C_{5V}\tfrac{1}{2} [ \Tr(\xi^{(n)}_{\nu}\Sigma) \Tr(\xi^{(n)}_{\nu}\Sigma^{\dagger})]\Bigr)\nonumber\\
&=-\frac{a^2}{8f^4}(C_{2V}+C_{5V})\,\sum_{\nu}\tau_{\nu ab}\tau_{\nu cd}\,\phi^a_{ij}\phi^b_{ji}\,\phi^c_{kl}\phi^d_{lk} + \dots \label{dplusV4pt}\\
-a^2\sum_\nu\Bigl(C_{2A}&\tfrac{1}{4} [ \Tr(\xi^{(n)}_{\nu5}\Sigma)\Tr(\xi^{(n)}_{5\nu}\Sigma) + h.c.]
+C_{5A}\tfrac{1}{2} [ \Tr(\xi^{(n)}_{\nu5}\Sigma) \Tr(\xi^{(n)}_{5\nu}\Sigma^{\dagger}) ]\Bigr)\nonumber\\
&=-\frac{a^2}{8f^4}(C_{2A}+C_{5A})\,\sum_{\nu}\tau_{\nu5, ab}\tau_{\nu5, cd}\,\phi^a_{ij}\phi^b_{ji}\,\phi^c_{kl}\phi^d_{lk} + \dots \label{dplusA4pt}
\end{align}
Finally, expanding the Gasser-Leutwyler Lagrangian of Eq.~(\ref{eq:GL}) through $\CO(\phi^2)$ gives
\begin{align}
L_4\Tr(\dmu\Sigma^\dagger\dmu\Sigma)\Tr(\chi^\dagger\Sigma+\chi\Sigma^\dagger)&=\frac{8L_4}{f^2}\dmu\phi_{ij}^a\dmu\phi_{ji}^a\,4(U_5+D_5+S_5)+\dots\label{eq:L4}\\
L_5\Tr(\dmu\Sigma^\dagger\dmu\Sigma(\chi^\dagger\Sigma+\Sigma^\dagger\chi))&=\frac{8L_5}{f^2}\dmu\phi_{ij}^a\dmu\phi_{ji}^a\,I_5+\dots\label{eq:L5}\\
-L_6[\Tr(\chi^\dagger\Sigma+\chi\Sigma^\dagger)]^2&=\frac{16L_6}{f^2}\phi_{ij}^a\phi_{ji}^a\,I_5\,4(U_5+D_5+S_5)+\dots\label{eq:L6}\\
-L_8\Tr(\chi^\dagger\Sigma\chi^\dagger\Sigma) + h.c.&=\frac{8L_8}{f^2}\phi_{ij}^a\phi_{ji}^a\,I_5\,(I_5+J_5)+\dots\label{eq:L8}
\end{align}
The normalization of Eqs.~(\ref{eq:L4}) and (\ref{eq:L6}) differs from
that in the continuum $\chi$PT by additional factors of $4$.  These
factors are systematically canceled by factors of $1/4$ when using
the replica method \cite{Damgaard:2000gh,Aubin:2003mg,Aubin:2003rg} to
arrive at the results for the 1+1+1 flavor theory.
\end{widetext}

\subsubsection{\label{sec:vres}Results by vertex class}
We consider external fields $\phi_{xy}^t$ and $\phi_{yx}^t$, where $t$ is the taste index, $x\neq y$, and we use the renormalization scheme of Refs.~\cite{Gasser:1984gg,Aubin:2003mg}.  For the tadpole graphs with kinetic energy vertices (Eq.~(\ref{KE4pt})), we find
\begin{eqnarray}
&&\frac{1}{12f^2}\sum_a\Biggl[p^2\Biggl(\sum_i (K_{xi,ix}^a + K_{yi,iy}^a) - 2\theta^{at}K_{xx,yy}^a\Biggr)\nonumber \\
&&\phantom{\frac{1}{12f^2}\sum_a\Biggl[}+\sum_i(L_{xi,ix}^a + L_{yi,iy}^a) - 2\theta^{at}L_{xx,yy}^a \Biggr],\label{KE444pre}
\end{eqnarray}
where $i=u,d,s$ runs over the flavors in the loops, $a$ is the taste
of mesons in the loops, $\theta^{ab}\equiv\frac{1}{4}\tau_{abab}=\pm1$
if $T^aT^b \mp T^bT^a = 0$, and
\begin{eqnarray}
K_{ij,kl}^a&\equiv&\int \frac{d^4q}{(2\pi)^4}\langle\phi_{ij}^a\phi_{kl}^a\rangle\label{Kintdef}\\
L_{ij,kl}^a&\equiv&\int \frac{d^4q}{(2\pi)^4}q^2\langle\phi_{ij}^a\phi_{kl}^a\rangle.
\end{eqnarray}
Substituting for the propagators and performing the integrals for the connected contributions gives
\begin{eqnarray}
&&\frac{1}{12f^2}\sum_a\Biggl[\frac{1}{(4\pi)^2}\sum_{Q}(p^2-Q_a)l(Q_a)\nonumber\\
&&\phantom{\frac{1}{12}}+\int\frac{d^4q}{(2\pi)^4}(p^2+q^2)(D_{xx}^a + D_{yy}^a - 2\theta^{at}D_{xy}^a)\Biggr],\label{KE444}
\end{eqnarray}
where $Q$ runs over the six flavor combinations $xi$, $yi$ for $i\in\{u,d,s\}$, $Q_a$ is the squared tree-level meson mass with flavor $Q$ and taste $a$, and
\begin{align}
l(X)\equiv X\Bigl(\ln X/\Lambda^2 + \delta_1(\sqrt{X}L)\Bigr)
\end{align}
for any squared meson mass $X$.
The finite-volume correction $\delta_1(\sqrt{X}L)$ is~\cite{Bernard:2001yj}
\begin{align}
\delta_1(\sqrt{X}L)\equiv
\frac{4}{\sqrt{X}L}\sum_{\vec{n}\ne\vec{0}}\frac{K_1(|\vec{n}|\sqrt{X}L)}{|\vec{n}|},
\end{align}
and $\delta_1(\sqrt{X}L)\to 0$ in infinite volume.  Here the temporal extent of the lattice is assumed infinite, $L$ is the spatial extent of the lattice, $\sqrt{X}L$ is assumed large for all values of the quark masses, $K_1$ is a Bessel function of imaginary argument, and the momentum in units of $2\pi/L$, $\vec{n}\in\mathbb{Z}^3$, is summed over all integer components except $\vec{n}=\vec{0}$.

The diagrams with the mass vertices (Eq.~(\ref{mass4pt})) are
\begin{eqnarray}
&&\frac{1}{12f^2}\sum_a\Biggl[\sum_i\Bigl[(m_{xy}^2+m_{xi}^2)K_{xi,ix}^a
+(m_{xy}^2+m_{yi}^2)K_{yi,iy}^a\Bigr]\nonumber\\
&&\phantom{\frac{1}{12f^2}\sum_a\Biggl[}+ 2m_{xy}^2\theta^{at}K_{xx,yy}^a\Biggr],\label{mass4ptloops}
\end{eqnarray}
where $m_{ij}^2=\mu(m_i+m_j)$.  Substituting for the propagators and performing the integrals for the connected diagrams gives
\begin{eqnarray}
&&\frac{1}{12f^2}\sum_a\Biggl[\frac{1}{(4\pi)^2}\sum_Q(P_5+Q_5)l(Q_a)\nonumber\\
&&+\int\frac{d^4q}{(2\pi)^4}\Bigl[(P_5+X_5)D_{xx}^a + (P_5+Y_5)D_{yy}^a\nonumber\\
&&\phantom{\frac{1}{12}\int\frac{d^4q}{(2\pi)^4}}+2P_5\theta^{at}D_{xy}^a\Bigr]\Biggr],\label{mass444}
\end{eqnarray}
where $X_5=m_{xx}^2$ and $Y_5=m_{yy}^2$.

For the graphs with vertices in Eqs.~(\ref{C1_4pt}), (\ref{C6_4pt}), (\ref{C3_4pt}), and (\ref{C4_4pt}), from the potential $\mathcal{U}$, we obtain
\begin{align}
\frac{a^2}{12f^2}\sum_a\Biggl[\Delta_{at}\sum_i(K_{xi,ix}^a+K_{yi,iy}^a)
+2\Delta^\prime_{at}K_{xx,yy}^a\Biggr],\label{C4ptloops}
\end{align}
where 
\begin{align}
\Delta_{at}&\equiv\frac{8}{f^2}\sum_{b\neq I} C_b(5+3\theta^{ab}\theta^{bt}-4\theta^{5b}\theta^{bt}-4\theta^{ab}\theta^{b5})\label{dat_def}\\
\Delta^\prime_{at}&\equiv\frac{8\theta^{at}}{f^2}\sum_{b\neq I} C_b(1+3\theta^{ab}\theta^{bt}-2\theta^{5b}\theta^{bt}-2\theta^{ab}\theta^{b5}),\label{eat_def}
\end{align}
and the sum over $b$ includes all tastes appearing in the taste matrices of the vertices from $\mathcal{U}$; {\it i.e.}, all tastes except the taste singlet.  The coefficients $C_b$ are the couplings of the vertices from $\mathcal{U}$:
\begin{eqnarray}
C_b=
\begin{cases}
C_1 & \text{if $b=5$}\\
C_6 & \text{if $b\in\{\mu\nu\}$}\\
C_3 & \text{if $b\in\{\mu\}$}\\
C_4 & \text{if $b\in\{\mu5\}$.}
\end{cases}
\end{eqnarray}
Substituting for the propagators and performing the integrals for the connected contributions gives
\begin{align}
&\frac{a^2}{12f^2}\sum_a\Biggl[\frac{\Delta_{at}}{(4\pi)^2}\sum_Q l(Q_a)\nonumber\\
&+\int\frac{d^4q}{(2\pi)^4}\Bigl[\Delta_{at}(D_{xx}^a+D_{yy}^a) + 2\Delta^\prime_{at}D_{xy}^a\Bigr]\Biggr].\label{U444}
\end{align}

For the contributions from the $\mathcal{U^\prime}$ (hairpin) vertices of Eqs.~(\ref{delV_4pt}) and (\ref{delA_4pt}), we have
\begin{equation}
\frac{1}{6f^2}\sum_{a\in V,A}\delta_a(2+\theta^{at})\sum_i(K_{ii,xx}^a+K_{ii,yy}^a),
\end{equation}
where $a$ runs over the taste vector and taste axial irreps, $V=\{\mu\}$ and $A=\{\mu5\}$.
Proceeding as above, this result becomes
\begin{align}
\frac{1}{6f^2}\sum_{a\in V,A}&\delta_a(2+\theta^{at})\Biggl[\frac{1}{(4\pi)^2}(l(X_a)+l(Y_a))\nonumber\\
&+\int\frac{d^4q}{(2\pi)^4}\sum_i\Bigl[D_{ix}^a+D_{iy}^a\Bigr]\Biggr],\label{hair444pre}
\end{align}
where $X_a$ and $Y_a$ are given by Eq.~\eqref{eq:fntm}.  The connected
and disconnected pieces of this result can be combined into a single
disconnected piece by using the identity~\cite{Aubin:2003mg}
\begin{equation}
\frac{1}{q^2+I_a}+\sum_j D_{ij}^a = -\frac{q^2+L_a}{\delta_a}D_{il}^a,\label{magic_id}
\end{equation}
where $I_a$ and $L_a$ are given by Eq.~\eqref{eq:fntm}, $i,j,l \in
\{u,d,s\}$, and $a\in \{V,A,I\}$.  This result follows immediately from the form of
$D_{ij}^a$ obtained by treating the $\CO(\phi^2)$ terms of the LO Lagrangian
that are proportional to $\delta_a$ as vertices and summing the resulting
geometric series for the flavor-neutral propagators:
\begin{equation}
D_{ij}^a=-\frac{\delta_a}{(q^2+I_a)(q^2+J_a)}\frac{1}{1+\delta_a{\displaystyle \sum_{l}}\dfrac{1}{q^2+L_a}}.
\end{equation}
The equivalence of this form and that given in Eq.~(\ref{D_piece}) was
demonstrated in Ref.~\cite{Sharpe:2000bc} for general partially quenched theories.  With the use of the
identity Eq.~(\ref{magic_id}), Eq.~(\ref{hair444pre}) becomes
\begin{equation}
-\frac{1}{3f^2}\sum_{a\in V,A}(2+\theta^{at})\int\frac{d^4q}{(2\pi)^4}(q^2+P_a)D_{xy}^a.\label{hair444}
\end{equation}

For the loops from the remaining two vertices of $\mathcal{U^\prime}$, 
Eqs.~(\ref{dplusV4pt}) and (\ref{dplusA4pt}), we find
\begin{equation}
\frac{a^2}{32f^2}\sum_a\Biggl(\sum_{b\in V,A}
\omega_b \tau_{abt}\tau_{abt}(1+\theta^{ab})\Biggr)K_{xy,yx}^a,
\end{equation}
where
\begin{align}
\label{eq:plus_hairpin}
\omega_b \equiv\frac{16}{f^2}
\begin{cases}
C_{2V}+C_{5V} & \text{if $b\in\{\mu\}$} \\
C_{2A}+C_{5A} & \text{if $b\in\{\mu5\}$.}
\end{cases}
\end{align}
Performing the integrals gives
\begin{equation}
\frac{a^2}{32f^2}\sum_a\Biggl(\sum_{b\in V,A}\omega_b\tau_{abt}\tau_{abt}(1+\theta^{ab})\Biggr)\frac{l(P_a)}{(4\pi)^2}.\label{hairplus444}
\end{equation}

For the tree-level graphs with the vertices in Eqs.~(\ref{eq:L4}), (\ref{eq:L5}), (\ref{eq:L6}), and (\ref{eq:L8}), we have
\begin{align}
&-\frac{16}{f^2}(2L_6\,P_5+L_4\,p^2)\,4(U_5+D_5+S_5)\label{eq:L6pL4}\\
&-\frac{16}{f^2}(2L_8\,P_5+L_5\,p^2)\,P_5,\label{eq:L8pL5}
\end{align}
while the tree-level graphs with vertices from the Sharpe-Van de Water Lagrangian may be parametrized by introducing LECs corresponding to the irreps of $SO(4)_T$ and $SW_4$:
\begin{align}
-\frac{16}{f^2}a^2(&\mc{A}_tP_5 + \mc{B}_t\,4(U_5+D_5+S_5)\nonumber\\
&+ \mc{C}_tp_4^2 + \mc{D}_ta^2),\label{eq:SV}
\end{align}
where the coefficients $\mc{A}_t$, $\mc{B}_t$, and $\mc{D}_t$ are degenerate
within the $SO(4)_T$ irreps, and the coefficients $\mc{C}_t$ are degenerate
within the $SW_4$ irreps.  The symmetry of the calculation under interchange of
the (valence) flavors $x\leftrightarrow y$ implies the $\CO(a^2m_q)$ terms are
proportional to $P_5$ or the sum $U_5+D_5+S_5$.
\subsection{\label{sec:res444}Results in 4+4+4 theory}
The results in Eqs.~(\ref{KE444}), (\ref{mass444}), (\ref{U444}),
(\ref{hair444}), and (\ref{hairplus444}) are the one-loop
contributions to the expansion of the (negative of the) self-energies
of the flavor-charged PGBs of taste
$t\in\{I,\mu,\mu\nu(\mu<\nu),\mu5,5\}$ in the 4+4+4 theory.
Collecting the connected contributions and factoring $-1/(4\pi f)^2$
gives
\begin{align}
\sigma^{con}&(p^2)=-\frac{1}{12}\sum_{a,Q}\biggl(p^2+P_5+a^2\Delta_{at}-a^2\Delta_a\biggr)l(Q_a)\nonumber\\
&-\frac{a^2}{32}\sum_a\Biggl[\sum_{b\in V,A}\omega_b\tau_{abt}\tau_{abt}(1+\theta^{ab})\Biggr]l(P_a).
\end{align}
where $Q$ runs over the six flavor combinations $xi$, $yi$ for $i\in\{u,d,s\}$,
$Q_a$ is the squared tree-level meson mass with flavor $Q$ and taste $a$, and
$P_a$ is given by Eq.~\eqref{eq:tm}.
Setting
$p^2=-P_t=-P_5-a^2\Delta_t$, we have
\begin{align}
\sigma^{con}&(-P_t)=-\frac{1}{12}
\sum_{a,Q}\biggl(a^2\Delta_{at}-a^2(\Delta_t+\Delta_a)
\biggr)l(Q_a)\nonumber\\
&-\frac{a^2}{32}\sum_a
\Biggl[\sum_{b\in V,A}\omega_b\tau_{abt}\tau_{abt}
(1+\theta^{ab})\Biggr]l(P_a).\label{con444}
\end{align}
For the Goldstone case, $t=5$ and Eq.~(\ref{dat_def}) with
Eqs.~(\ref{Itsplitdef}) through (\ref{Ptsplitdef}) imply
\begin{equation}
\Delta_{a5}=\Delta_a,\label{reduce}
\end{equation}
while
\begin{equation}
\tau_{ab5}\tau_{ab5}(1+\theta^{ab})=0\ \text{if $b\in V,A$,}
\end{equation}
so the connected contributions vanish identically~\cite{Aubin:2003mg}.  

The chiral logarithms are degenerate within the $SO(4)_T$ irreps; summing over the values of $a$ within each irrep, we rewrite Eq.~(\ref{con444}):
\begin{equation}
\sigma^{con}=-a^2\sum_B
\Biggl(\delta_{BF}^{con}\,l(P_B)+ 
\frac{\Delta_{BF}^{con}}{12}\sum_Q l(Q_B)\Biggr),\label{fin_con444}
\end{equation}
where
\begin{align}
\delta_{BF}^{con}&\equiv
\frac{1}{32}\sum_{a\in B}\sum_{b\in V,A}
\omega_b\tau_{abt}\tau_{abt}(1+\theta^{ab})\label{delBFcon_def}\\
\Delta_{BF}^{con}&\equiv\sum_{a\in B}\biggl(\Delta_{at}-(\Delta_t+\Delta_a)\biggr).\label{DelBFcon_def}
\end{align}
The indices $B$ and $F$ designate the $SO(4)_T$ irreps, 
$B,F\in\{I,V,T,A,P\}$, 
$t\in F$, and we conveniently abuse the notation to define
\begin{align}
X_B\equiv X_a\quad\text{for $a\in B$ and $X\in\{P,Q\}$},
\end{align}
which is possible because the taste splittings are degenerate within irreps of $SO(4)_T$.
\begin{table}[htbp]
\begin{center}
\caption{\label{ConDiag2}The coefficients $\delta_{BF}^{con}$ defined in Eq.~(\ref{delBFcon_def}).  To obtain $\delta_{BF}^{con}$, multiply the entry in row $B$ and column $F$ by $16/f^2$.}
\begin{tabular}{c | c | c}
\hline\hline
$B \backslash F$ & $V$ & $A$\\
\hline
$V$ & 0 & 0 \\
$A$ & 0 & 0 \\
$T$ & $3(C_{2A}+C_{5A})$ & $3(C_{2V}+C_{5V})$\\
$P$ & 0 & 0 \\
$I$ & $C_{2V}+C_{5V}$ & $C_{2A}+C_{5A}$\\
\hline\hline
\end{tabular}
\phantom{dummycharlonglonglonglongsdfsdfsdfsfsfsfsdf}\\
\begin{tabular}{c | c | c | c }
\hline\hline
$B \backslash F$ & $T$ & $P$ & $I$ \\
\hline
$V$ & $2(C_{2A}+C_{5A})$ & 0 & $4(C_{2V}+C_{5V})$\\
$A$ & $2(C_{2V}+C_{5V})$ & 0 & $4(C_{2A}+C_{5A})$\\
$T$ & 0 & 0 & 0 \\
$P$ & 0 & 0 & 0 \\
$I$ & 0 & 0 & 0 \\
\hline\hline
\end{tabular}
\end{center}
\end{table}
\color{black}
\begin{table*}[htbp]
\caption{\label{ConDiag}The coefficients $\Delta_{BF}^{con}$ defined 
  in Eq.~(\ref{DelBFcon_def}).  To obtain $\Delta_{BF}^{con}$, 
  multiply the entry in row $B$ and column $F$ by $96/f^2$.}
\begin{center}
\begin{tabular}{c | c | c | c | c | c }
\hline\hline
$B \backslash F$ & $V$ & $A$ & $T$ & $P$ & $I$ \\
\hline
$V$ & $4C_1+C_3+9C_4+6C_6$ & $4C_1+3C_3+3C_4+6C_6$ 
  & $2C_3+6C_4+8C_6$ & $0$ & $4C_3+12C_4$ \\
$A$ & $4C_1+3C_3+3C_4+6C_6$ & $4C_1+9C_3+C_4+6C_6$
  & $6C_3+2C_4+8C_6$ & $0$ & $12C_3+4C_4$\\
$T$ & $3C_3+9C_4+12C_6$ & $9C_3+3C_4+12C_6$
  & $6C_3+6C_4+16C_6$ & $0$ & $12C_3+12C_4$\\
$P$ & $0$ & $0$
  & $0$ & $0$ & $0$\\
$I$ & $C_3+3C_4$ & $3C_3+C_4$
  & $2C_3+2C_4$ & $0$ & $4C_3+4C_4$\\
\hline\hline
\end{tabular}
\end{center}
\end{table*}

The coefficients $\delta_{BF}^{con}$ and $\Delta_{BF}^{con}$ are linear combinations of the couplings in the potentials $\mathcal{U^\prime}$ and $\mathcal{U}$, respectively.  Equivalently, the coefficients $\Delta_{BF}^{con}$ are linear combinations of the taste splittings $\Delta_a$, and the coefficients $\delta_{BF}^{con}$ are linear combinations of the couplings $\omega_{V,A}$ (defined in Eq.~(\ref{eq:plus_hairpin})).

Explicit results for $\delta_{BF}^{con}$ and $\Delta_{BF}^{con}$ are
given in Tables~\ref{ConDiag2} and \ref{ConDiag}.  We note that
$\delta_{PF}^{con}=\Delta_{PF}^{con}=0$.  $\Delta_{PF}^{con}=0$
follows from the symmetry of the summand of Eq.~(\ref{DelBFcon_def})
under interchange of the indices $a,t$ and the vanishing of the
connected contributions in the Goldstone case, $\Delta_{BP}^{con}=0$.
The symmetry $a\leftrightarrow t$ does not exist in the summand of
Eq.~(\ref{delBFcon_def}), so the relation $\delta_{PF}^{con}=0$
appears non-trivial.
In general
the symmetry $a\leftrightarrow t$ of the summand of
Eq.~(\ref{DelBFcon_def}) implies
\begin{align}
N_F\Delta_{BF}^{con}=N_B\Delta_{FB}^{con},\label{checker}
\end{align}
where $N_{F(B)}$ is the dimension of irrep $F(B)$.
Eq.~(\ref{checker}) is useful for checking the results in Table~\ref{ConDiag}.
Eq.~(\ref{fin_con444}), with the coefficients in Tables~\ref{ConDiag2} and \ref{ConDiag}, is our final result for the connected tadpoles in the 4+4+4 theory.

Collecting the disconnected pieces from the one-loop results in Eqs.~(\ref{KE444}), (\ref{mass444}), (\ref{U444}), (\ref{hair444}), and (\ref{hairplus444}) gives
\begin{align}
\sigma^{disc}(p^2)=&-\frac{(4\pi)^2}{12}\sum_a\int\frac{d^4q}{(2\pi)^4}\Biggl[\biggl(p^2+P_5\nonumber\\
&+a^2\Delta_{at}-a^2\Delta_a\biggr)(D_{xx}^a+D_{yy}^a)\nonumber\\
&+\biggl[-2\theta^{at}p^2 + \Bigl(2(1-\theta^{at}) + \rho^{at}\Bigr)q^2\nonumber\\
&+ \Bigl(2(1+\theta^{at}) + \rho^{at}\Bigr)P_5\nonumber\\
&+ 2a^2\Delta^\prime_{at} + (2+\rho^{at})a^2\Delta_a\biggr]D_{xy}^a\Biggr],
\end{align}
where
\begin{align}
\rho^{at}\equiv
\begin{cases}
 -4(2+\theta^{at}) & \text{if $a\neq I$} \\
 0 & \text{if $a=I$.}
\end{cases}
\label{eq:rho}
\end{align}
Setting $p^2=-P_t=-P_5-a^2\Delta_t$, we find
\begin{align}
\sigma&^{disc}(-P_t)=-\frac{(4\pi)^2}{12}\sum_a\int\frac{d^4q}{(2\pi)^4}\Biggl[\biggl(a^2\Delta_{at}\nonumber\\
&\phantom{^{disc}(-m_{xyt}^2)=}
-a^2(\Delta_t+\Delta_a)\biggr)(D_{xx}^a+D_{yy}^a)\nonumber\\
&+\biggl[\Bigl(2(1-\theta^{at}) + \rho^{at}\Bigr)q^2
+\Bigl(2(1+2\theta^{at}) + \rho^{at}\Bigr)P_5\nonumber\\
&+ 2a^2\Delta^\prime_{at} + a^2\Bigl(2\theta^{at}\Delta_t + (2+\rho^{at})\Delta_a\Bigr)\biggr]D_{xy}^a\Biggr].\label{disc444pre}
\end{align}
For the Goldstone case, $t=5$ and Eqs.~(\ref{dat_def}), (\ref{eat_def}), and (\ref{reduce}) imply
\begin{equation}
\Delta^\prime_{a5}=-\theta^{a5}\Delta_a,
\end{equation}
and Eq.~(\ref{disc444pre}) reduces to the result of Ref.~\cite{Aubin:2003mg}.

The sum over $a$ receives nonzero contributions from the $SO(4)_T$
vector, axial, and singlet irreps.  Summing over $a$ within each
irrep, we can write
\begin{align}
\sigma&^{disc}=-\frac{(4\pi)^2}{12}\int\frac{d^4q}{(2\pi)^4}\Biggl[a^2\Delta_{VF}^{con}
(D_{xx}^V+D_{yy}^V)\nonumber\\
&+2\biggl(-12P_5
-6q^2\nu_{VF}+a^2\Delta_{VF}^{disc}\biggr)D_{xy}^V\nonumber\\
&+(V\rightarrow A)
+a^2\Delta_{IF}^{con}(D_{xx}^I+D_{yy}^I)\nonumber\\
&+2(3P_5+a^2\Delta_{IF}^{disc})D_{xy}^I
\Biggr],\label{disc444}
\end{align}
where
\begin{align}
\Delta_{BF}^{disc}&\equiv\sum_{a\in B}\biggl(\Delta^\prime_{at}+\theta^{at}\Delta_t + (1+\rho^{at}/2)\Delta_a\biggr)\label{DeltaDBFDef}\\
\nu_{BF}&\equiv\frac{1}{2}\sum_{a\in B}(1+\theta^{at})
\end{align}
for $t\in F$.  $\nu_{BF}$ is the number of taste matrices for irrep $B\in\{V,A\}$ that commute with the taste matrix corresponding to $t\in F$.  The values of $\nu_{BF}$ are given in Table~\ref{tab:numcom}.
\begin{table}[h]
\begin{center}
\caption{\label{DiscDiag}The coefficients $\Delta_{BF}^{disc}$ defined in Eq.~(\ref{DeltaDBFDef}).  To obtain $\Delta_{BF}^{disc}$, multiply the entry in row $B$ and column $F$ by $96/f^2$.}
\begin{tabular}{c | c | c }
\hline\hline
$B \backslash F$ & $V$ & $A$\\
\hline
$V$ & $-3C_1-6C_4-3C_6$ & $-C_1-6C_4-9C_6$\\
$A$ & $-C_1-6C_3-9C_6$ & $-3C_1-6C_3-3C_6$\\
$I$ & $C_3+3C_4$ & $3C_3+C_4$\\
\hline\hline
\end{tabular}
\phantom{dummychadfgdfgdfgdfgdfgdfgdfgdgdfgr}\\
\begin{tabular}{c | c | c | c }
\hline\hline
$B \backslash F$ & $T$ & $P$ & $I$\\
\hline
$V$ & $-2C_1-8C_4-6C_6$ & $0$ & $-4C_1-12C_6$\\
$A$ & $-2C_1-8C_3-6C_6$ & $0$ & $-4C_1-12C_6$\\
$I$ & $2C_3+2C_4$ & $0$ & $4C_3+4C_4$\\
\hline\hline
\end{tabular}
\end{center}
\end{table}
\begin{table}[h]
\begin{center}
\caption{\label{tab:numcom}The numbers $\nu_{BF}$ of taste matrices for irrep $B\in\{V,A\}$ that commute with any given taste matrix for irrep $F$.  $\nu_{BF}$ appears in row $B$ and column $F$.}
\begin{tabular}{c | c | c | c | c | c}
\hline\hline
$B \backslash F$ & $V$ & $A$ & $T$ & $P$ & $I$\\
\hline
               $V$ & 1 & 3 & 2 & 0 & 4\\ 
               $A$ & 3 & 1 & 2 & 0 & 4\\ 
\hline\hline
\end{tabular}
\end{center}
\end{table}

The coefficients $\Delta_{BF}^{disc}$, like the coefficients $\Delta_{BF}^{con}$, are linear combinations of the taste splittings $\Delta_a$.  In Appendix~\ref{sum_theta} we show that
\begin{align}
\Delta_{BP}^{disc}&=0\quad\text{for $B=I,V,A$}\\
\Delta_{VI}^{disc}&=\Delta_{VI}^{con}-24\Delta_V\label{eq:DelVIdisc_chk}\\
\Delta_{AI}^{disc}&=\Delta_{AI}^{con}-24\Delta_A\label{eq:DelAIdisc_chk}.
\end{align}
The latter two relations follow from the identity
\begin{equation}
\Delta^\prime_{at}=\Delta_{at}-2(\Delta_a+\Delta_t)\quad \text{if $\theta^{at}=1$.}
\label{EatIdentity}
\end{equation}
Eqs.~\eqref{eq:DelVIdisc_chk} and \eqref{eq:DelAIdisc_chk} provide non-trivial checks of the results for $\Delta_{BI}^{disc}$ in Table~\ref{DiscDiag}.  Eq.~(\ref{disc444}), with the coefficients in Tables~\ref{ConDiag}, \ref{DiscDiag}, and \ref{tab:numcom}, is our final result for the disconnected tadpoles in the 4+4+4 theory.  

Taking into account the hairpin couplings, taste splittings, and coefficients $\delta_{BF}^{con}$, $\Delta_{BF}^{con}$, and $\Delta_{BF}^{disc}$ in Tables~\ref{ConDiag2}, \ref{ConDiag}, and \ref{DiscDiag}, we see that the loops, the results in Eqs.~(\ref{fin_con444}) and (\ref{disc444}), are invariant under the symmetry
\begin{align}
C_3&\leftrightarrow C_4\\
C_{2V}&\leftrightarrow C_{2A}\\
C_{5V}&\leftrightarrow C_{5A}
\end{align}
or, more briefly, under
\begin{equation}
V\leftrightarrow A
\end{equation}
in accord with the observations following Eq.~(\ref{eq:transf}) above.

Collecting the analytic contributions to the self-energies from Eqs.~(\ref{eq:L6pL4}), (\ref{eq:L8pL5}), and (\ref{eq:SV}) gives
\begin{align}
\sigma^{anal}(p_4^2)&=
\frac{16}{f^2}(2L_6\,P_5+L_4\,p_4^2)\,4(U_5+D_5+S_5)\nonumber\\
&+\frac{16}{f^2}(2L_8\,P_5+L_5\,p_4^2)\,P_5\nonumber\\
&+\frac{16}{f^2}a^2(\mc{A}_tP_5 + \mc{B}_t\,4(U_5+D_5+S_5)\nonumber\\
&\phantom{+\frac{16}{f^2}a^2(}+ \mc{C}_tp_4^2 + \mc{D}_ta^2).
\end{align}
Setting $p_4^2=-P_t=-P_5-a^2\Delta_t$, we have
\begin{align}
\sigma^{anal}=
&\phantom{+}\frac{16}{f^2}(2L_6-L_4)P_5\,4(U_5+D_5+S_5)\nonumber\\
&+\frac{16}{f^2}(2L_8-L_5)\,P_5^2\nonumber\\
&+\frac{16}{f^2}a^2(\mc{E}_tP_5 + \mc{F}_t\,4(U_5+D_5+S_5)\nonumber\\
&\phantom{+\frac{16}{f^2}a^2(}+ \mc{G}_ta^2),\label{eq:sanal444}
\end{align}
where we have absorbed terms proportional to $a^2\Delta_t$ into the coefficients $\mc{E}_t$, $\mc{F}_t$, and $\mc{G}_t$.
The first two lines of Eq.~\eqref{eq:sanal444} correspond to the continuum result and are the same for all tastes.  In the last two lines, the coefficients $\mc{F}_t$ are degenerate within irreps of $SO(4)_T$, while the coefficients $\mc{E}_t$ and $\mc{G}_t$ are degenerate within irreps of $SW_4$.  The exact chiral symmetry implies that $\mc{F}_5=\mc{G}_5=0$.  Setting $t=5$ in Eq.~\eqref{eq:sanal444} then yields the result of Ref.~\cite{Aubin:2003mg}.  In Appendix~\ref{app:pattern} we recall the results for the dispersion relations of Sharpe and Van de Water~\cite{Sharpe:2004is}; the $SO(4)_T$-breaking contributions to the $\mc{E}_t$ and $\mc{G}_t$ terms in Eq.~\eqref{eq:sanal444} come from only three operators in the Sharpe-Van de Water Lagrangian.  The consequences for fitting strategy are discussed in Sec.~\ref{sec:disc}.

Eqs.~\eqref{fin_con444}, \eqref{disc444}, and \eqref{eq:sanal444} are useful starting points for deriving results in various cases of interest.  In Sec.~\ref{sec:res} we use them to deduce results for fully dynamical, partially quenched, and quenched theories.

\section{\label{sec:res}Results}
The results in Eqs.~\eqref{fin_con444}, \eqref{disc444}, and
\eqref{eq:sanal444} must be modified to account for (partial)
quenching~\cite{Bernard:1992mk,Bernard:1993sv} and the fourth root of the staggered fermion determinant~\cite{Bazavov:2009bb}.  The replica method of Ref.~\cite{Damgaard:2000gh}
allows us to generalize to the partially quenched case.  The replica method
also allows us to systematically take into account the fourth root of the
staggered
determinant~\cite{Aubin:2003rg,Bernard:2006zw,Bernard:2006qt,Bernard:2007yt,Bernard:2007ma}.

The effects of partial quenching and rooting in Eqs.~(\ref{fin_con444}), (\ref{disc444}), and (\ref{eq:sanal444}) are easily summarized:  The valence quark masses $m_x$ and $m_y$ are no longer degenerate with the sea quark masses $m_u$, $m_d$, and $m_s$, a factor of $1/4$ is introduced in the second term of Eq.~(\ref{fin_con444}), the eigenvalues of the mass matrix in Eq.~(\ref{444_mass_matrix}) are replaced with the eigenvalues of
\begin{equation}\label{111_mass_matrix}
	\begin{pmatrix}
	U_a +\delta_a/4 & \delta_a/4 & \delta_a/4 \\
	\delta_a/4 & D_a +\delta_a/4 & \delta_a/4 \\
	\delta_a/4 & \delta_a/4 & S_a +\delta_a/4
	\end{pmatrix},
\end{equation}
and terms in Eq.~\eqref{eq:sanal444} that are proportional to the sum of the sea quark masses are multiplied by $1/4$.

Accounting for quenching the sea quarks is equally straightforward~\cite{Bernard:1992mk,Bernard:2001yj,Aubin:2003mg}:  The second term of Eq.~\eqref{fin_con444} is dropped, and the disconnected part of the propagator, in Eq.~\eqref{D_piece}, is everywhere replaced with
\begin{equation}
D^{a,\mathrm{quench}}_{il}=-\frac{\delta_a^\mathrm{quench}}{(q^2+I_a)(q^2+L_a)},\label{Dquench}
\end{equation}
where
\begin{equation}
\delta_a^\mathrm{quench}=
\begin{cases}
4(m_0^2+\alpha q^2)/3 & \text{if $a=I$}\\
\delta_a & \text{if $a\neq I$.}
\end{cases}\label{dquench}
\end{equation}

The one-loop contributions to the self-energies in the fully dynamical 1+1+1 and 2+1 flavor cases in the chiral $SU(3)$ theory are in Sec.~\ref{sec:fullsu3qcd}.  One-loop contributions for partially quenched 1+1+1 and 2+1 flavor cases of interest are in Sec.~\ref{sec:PQsu3qcd}.
In Secs.~\ref{sec:fullsu2qcd} and \ref{sec:PQsu2qcd} we write down the analogous (fully dynamical and partially quenched) one-loop contributions in the chiral $SU(2)$ theory.  Sec.~\ref{sec:Qsu3qcd} contains one-loop contributions for the quenched case.
%
%

\subsection{$SU(3)$ chiral perturbation theory}
\subsubsection{\label{sec:fullsu3qcd}Fully dynamical case}
Introducing a factor of $1/4$ in the second term of
Eq.~\eqref{fin_con444}, we obtain the connected loop contributions in the fully
dynamical 1+1+1 flavor case,
\begin{equation}
\sigma^{con}=-a^2\sum_B
\Biggl(\delta_{BF}^{con}\,l(P_B)+ 
\frac{\Delta_{BF}^{con}}{48}\sum_Q l(Q_B)\Biggr),
\label{phi_full_con111_su3}
\end{equation}
where $P_B$ is the LO squared mass $m_\phi^2$ of a flavor-charged PGB $\phi_{xy}^B$ with $\bar{x}y$ valence (anti)quarks ($x\ne y$), and
$Q_B$ is the LO squared mass $m_\phi^2$ of a PGB $\phi_{z\ell}^B$ with $\bar{z}\ell$ valence
(anti)quarks, where $z\in \{x,y\}$, $\ell \in \{u,d,s\}$, and the sum over $Q$ runs over the six flavor combinations formed by pairing the possibilities for $z$ with those for $\ell$.
Setting $xy=ud,us,ds$ gives the results for the $\pi^+$, $K^+$, and $K^0$.
We have
\begin{align}
\sigma&^{con}_{\pi^+} = -a^2\sum_B\Biggl(\delta_{BF}^{con}\,l(\pi^+_B) +
\frac{\Delta_{BF}^{con}}{48}\Bigl(l(U_B) \nonumber\\
&+ 2l(\pi^+_B) + l(K^+_B) + l(D_B) + l(K^0_B)\Bigr)\Biggr)\label{piplus_full_con111_su3}\\
\sigma&^{con}_{K^+} = -a^2\sum_B\Biggl(\delta_{BF}^{con}\,l(K^+_B) +
\frac{\Delta_{BF}^{con}}{48}\Bigl(l(U_B)\nonumber\\
&+ l(\pi^+_B) + 2l(K^+_B) + l(K^0_B) + l(S_B)\Bigr)\Biggr)\label{cK_full_con111_su3}\\
\sigma&^{con}_{K^0} = -a^2\sum_B\Biggl(\delta_{BF}^{con}\,l(K^0_B) +
\frac{\Delta_{BF}^{con}}{48}\Bigl(l(\pi^+_B)\nonumber\\
&+ l(D_B) + 2l(K^0_B) + l(K^+_B) + l(S_B)\Bigr)\Biggr),\label{nK_full_con111_su3}
\end{align}
where the squared tree-level masses of the flavor-charged mesons are
\begin{align}
\pi_B^+ &= \mu(m_u+m_d) + a^2\Delta_B\\
K_B^+ &= \mu(m_u+m_s) + a^2\Delta_B\\
K_B^0 &= \mu(m_d+m_s) + a^2\Delta_B.
\end{align}

The integrals of the disconnected pieces, in Eq.~\eqref{disc444}, can be performed by using the residues of Ref.~\cite{Aubin:2003mg} to expand the integrands.  For the $\pi^+$ we find
\begin{align}
\sigma&^{disc}_{\pi^+}=\frac{1}{12}\Biggl[a^4\Delta_{VF}^{con}\delta_V^\prime
\sum_X\biggl(R_{U\pi^0\eta\eta^\prime}^{DS}(X_V)l(X_V) \nonumber\\
&+ R_{D\pi^0\eta\eta^\prime}^{US}(X_V)l(X_V)\biggr)
\nonumber\\
&+2\sum_X\biggl(-12\pi_5^{+}+6X_V\nu_{VF}+a^2\Delta_{VF}^{disc}\biggr)\times \nonumber\\
&\mbox{\quad}a^2\delta_V^\prime R_{\pi^0\eta\eta^\prime}^{S}(X_V)l(X_V)+(V\rightarrow A) \nonumber\\
&+\frac{4}{3}a^2\Delta_{IF}^{con}\sum_{X}\biggl(R_{U\pi^0\eta}^{DS}(X_I)l(X_I) + R_{D\pi^0\eta}^{US}(X_I)l(X_I)\biggr) \nonumber\\
&+\frac{8}{3}(3\pi_5^{+}+a^2\Delta_{IF}^{con})\sum_{X}R_{\pi^0\eta}^{S}(X_I)l(X_I)\Biggr],
\label{pi_full_disc111_su3}
\end{align}
where we decoupled the flavor-taste singlet, $\eta_I^\prime$, in the taste singlet channel by taking $m_0^2\to \infty$ before expanding the integrands.  

In Eq.~\eqref{pi_full_disc111_su3} we introduce a few naively perverse but
extremely convenient abuses of notation.  First, in each sum over $X$, the
residue in the summand determines the values of the index $X$.  For example, in
\begin{align}
\sum_XR_{U\pi^0\eta\eta^\prime}^{DS}(X)l(X),
\end{align}
the index $X$ takes the values in the set $\{U,\pi^0,\eta,\eta^\prime\}$.  When
the summation over $X$ is factored from sums of residues specifying different
sets of values for the index $X$, as in the first and second lines of
Eq.~\eqref{pi_full_disc111_su3}, we first distribute the summation symbol and
then use the residues to specify the values of $X$ in each sum.  Second, the
$SO(4)_T$ irrep specified in the argument of a given residue applies to all
masses appearing in the residue.  For example,
\begin{align}
R_{U\pi^0\eta\eta^\prime}^{DS}(U_V)=\frac{(D_V-U_V)(S_V-U_V)}{(\pi^0_V-U_V)(\eta_V-U_V)(\eta^\prime_V-U_V)},
\end{align}
where we continue denoting squared tree-level masses by the names of the mesons.
In general the residues are
\begin{align}
R_{B_1B_2\cdots B_n}^{A_1A_2\cdots A_k}(X_F)\equiv\frac{\prod_{A_j}(A_{jF}-X_F)}{\prod_{B_{i}\ne X}(B_{iF}-X_F)},
\end{align}
where $X\in\{B_1,B_2,\dots,B_n\}$ and $F\in\{V,A,I\}$ is the $SO(4)_T$ irrep.
%
%
%

The results for the $K^+$ and $K^0$ may be
obtained by permuting $U,D,S$ in the residues of
Eq.~(\ref{pi_full_disc111_su3}) and replacing $\pi^+$ with $K^+,K^0$.  They are
\begin{align}
\sigma^{disc}_{K^+}&=\frac{1}{12}\sum_X\Biggl[a^4\Delta_{VF}^{con}\delta_V^\prime
\biggl(R_{U\pi^0\eta\eta^\prime}^{DS}(X_V)l(X_V)\nonumber\\
&+ R_{S\pi^0\eta\eta^\prime}^{UD}(X_V)l(X_V)\biggr) \nonumber\\
&+2\biggl(-12K_5^{+}+6X_V\nu_{VF}+a^2\Delta_{VF}^{disc}\biggr)\times \nonumber\\
&\quad a^2\delta_V^\prime R_{\pi^0\eta\eta^\prime}^{D}(X_V)l(X_V)+(V\rightarrow A) \nonumber\\
&+\frac{4}{3}a^2\Delta_{IF}^{con}\biggl(R_{U\pi^0\eta}^{DS}(X_I)l(X_I) + R_{S\pi^0\eta}^{UD}(X_I)l(X_I)\biggr) \nonumber\\
&+\frac{8}{3}(3K^{+}+a^2\Delta_{IF}^{con})R_{\pi^0\eta}^{D}(X_I)l(X_I)\Biggr]
\label{cK_full_disc111_su3}
\end{align}
%
%
and
\begin{align}
\sigma^{disc}_{K^0}&=\frac{1}{12}\sum_X\Biggl[a^4\Delta_{VF}^{con}\delta_V^\prime
\biggl(R_{D\pi^0\eta\eta^\prime}^{US}(X_V)l(X_V)\nonumber\\
&+ R_{S\pi^0\eta\eta^\prime}^{UD}(X_V)l(X_V)\biggr) \nonumber\\
&+2\biggl(-12K_5^{0}+6X_V\nu_{VF}+a^2\Delta_{VF}^{disc}\biggr)\times \nonumber\\
&\quad a^2\delta_V^\prime R_{\pi^0\eta\eta^\prime}^{U}(X_V)l(X_V)+(V\rightarrow A) \nonumber\\
&+\frac{4}{3}a^2\Delta_{IF}^{con}\biggl(R_{D\pi^0\eta}^{US}(X_I)l(X_I) + R_{S\pi^0\eta}^{UD}(X_I)l(X_I)\biggr) \nonumber\\
&+\frac{8}{3}(3K^{0}+a^2\Delta_{IF}^{con})R_{\pi^0\eta}^{U}(X_I)l(X_I)\Biggr].
\label{nK_full_disc111_su3}
\end{align}
In Eqs.~(\ref{pi_full_disc111_su3}), (\ref{cK_full_disc111_su3}), and (\ref{nK_full_disc111_su3}), the masses $\pi^0_B$, $\eta_B$, and $\eta^\prime_B$ are the eigenvalues of the mass matrix in Eq.~(\ref{111_mass_matrix}) with $m_u\neq m_d$.

For physical values of the quark masses, strong isospin breaking is a small correction to electromagnetic isospin breaking, and 2+1 flavor simulations have proven very useful~\cite{Bazavov:2009bb}.  
%
%
%
Setting $xy=ud,us$ and $m_u=m_d$ in Eq.~\eqref{phi_full_con111_su3} gives the connected contributions for $\pi$, $K$:
\begin{align}
\sigma^{con}_{\pi} =& -a^2\sum_B\Biggl(\delta_{BF}^{con}\,l(\pi_B) +
\frac{\Delta_{BF}^{con}}{24}\Bigl(2l(\pi_B) + l(K_B)\Bigr)\Biggr)\label{pi_full_con2p1_su3} \\
\sigma^{con}_{K} =& -a^2\sum_B\Biggl(\delta_{BF}^{con}\,l(K_B) +
\frac{\Delta_{BF}^{con}}{48}\Bigl(2l(\pi_B) + 3l(K_B) \nonumber\\
&\phantom{-a^2\sum_B\Biggl(\delta_{BF}^{con}\,l(K_B)}+ l(S_B)\Bigr)\Biggr).\label{K_full_con2p1_su3}
\end{align}

In the 2+1 flavor case, the disconnected contributions are most easily obtained by returning to Eq.~(\ref{disc444}) and performing the integrals after setting $m_u=m_d$.  We find
\begin{align}
\sigma^{disc}_{\pi}&=\frac{1}{12}\Biggl[2\sum_{X}\biggl(-12\pi_5+6X_V\nu_{VF} \nonumber\\
&+ a^2(\Delta_{VF}^{con} + \Delta_{VF}^{disc})\biggr)a^2\delta_V^\prime R_{\pi\eta\eta^\prime}^{S}(X_V) l(X_V)\nonumber\\
&+ (V\rightarrow A) \nonumber\\ 
&+\frac{8}{3}(3\pi_5+2a^2\Delta_{IF}^{con})\biggl(\frac{3}{2}l(\pi_I)-\frac{1}{2}l(\eta_I)\biggr)\Biggr],
\label{isopi_full_disc2p1_su3}
\end{align}
\begin{align}
\sigma^{disc}_{K}&=\frac{1}{12}\Biggl[a^4\Delta_{VF}^{con}\delta_V^\prime\sum_X
\biggl(R_{\pi\eta\eta^\prime}^{S}(X_V)l(X_V)\nonumber\\
&+ R_{S\eta\eta^\prime}^{\pi}(X_V)l(X_V)\biggr)\nonumber\\
&+2\sum_X\biggl(-12K_5+6X_V\nu_{VF}+a^2\Delta_{VF}^{disc}\biggr)\times \nonumber\\
&\mbox{\quad}a^2\delta_V^\prime R_{\eta\eta^\prime}(X_V) l(X_V)+(V\rightarrow A) \nonumber\\ 
&+4a^2\Delta_{IF}^{con}\left(\frac{1}{2}l(\pi_I)+l(S_I)\right) 
\nonumber \\
&+\frac{8}{3}\left(3K_5-\frac{1}{4}a^2\Delta_{IF}^{con}\right)l(\eta_I)\Biggr],
\label{K_full_disc2p1_su3}
\end{align}
where $\pi_B$, $\eta_B$, and $\eta^\prime_B$ ($B\in\{V,A,I\}$) are the
eigenvalues of the mass matrix in Eq.~\eqref{111_mass_matrix} with $m_u=m_d$,
and we used the relations of the tree-level masses in the taste singlet channel
to simplify the associated residues:
\begin{align}
R_{\pi\eta}^{S}(\pi_I)=\frac{3}{2}&\quad
R_{\pi\eta}^S(\eta_I)=-\frac{1}{2}\\
R_{S\eta}^\pi(S_I)=3&\quad
R_{S\eta}^\pi(\eta_I)=-2.
\end{align}
%
%
%
In the continuum limit, only the taste singlet contributions to the
disconnected loops survive.  Taking the continuum limits of
Eqs.~\eqref{pi_full_con2p1_su3} through \eqref{K_full_disc2p1_su3}, we recover
the one-loop results of Gasser and Leutwyler~\cite{Gasser:1984gg}.  

\subsubsection{\label{sec:PQsu3qcd}Partially quenched case}
The connected contributions in the partially quenched 1+1+1 flavor
case have the same form as the connected contributions in the fully
dynamical 1+1+1 flavor case, Eq.~\eqref{phi_full_con111_su3}.  The
difference is that the valence and sea quark masses are, in general, non-degenerate:
$m_x,m_y\notin \{m_u,m_d,m_s\}$.
%
%

For the disconnected contributions in the 1+1+1 flavor case, keeping all quark masses in Eq.~\eqref{disc444} distinct and performing the integrals as before, we find
%
%
\begin{align}
\sigma&^{disc}_{x\ne y}=\frac{1}{12}\Biggl[a^4\Delta_{VF}^{con}\delta_V^\prime
\biggr(R_{X\pi^0\eta\eta^\prime}^{UDS}(X_V) \tilde{l}(X_V)\nonumber\\
&+ R_{Y\pi^0\eta\eta^\prime}^{UDS}(Y_V)\tilde{l}(Y_V) + \sum_Z\Bigl(D_{X\pi^0\eta\eta^\prime,X}^{UDS}(Z_V)l(Z_V)\nonumber\\
&+ D_{Y\pi^0\eta\eta^\prime,Y}^{UDS}(Z_V)l(Z_V)\Bigr)\biggl)\nonumber\\
&+2\sum_Z\biggl(-12P_5+6Z_V\nu_{VF}+a^2\Delta_{VF}^{disc}\biggr)\times \nonumber\\
&\quad a^2\delta_V^\prime R_{XY\pi^0\eta\eta^\prime}^{UDS}(Z_V)l(Z_V)+(V\rightarrow A) \nonumber\\ 
&+\frac{4}{3}a^2\Delta_{IF}^{con}
\biggl(R_{X\pi^0\eta}^{UDS}(X_I)\tilde{l}(X_I)
+ R_{Y\pi^0\eta}^{UDS}(Y_I)\tilde{l}(Y_I) \nonumber\\
&+ \sum_Z\Bigl(D_{X\pi^0\eta,X}^{UDS}(Z_I)l(Z_I) 
+ D_{Y\pi^0\eta,Y}^{UDS}(Z_I)l(Z_I)\Bigr)\biggr) \nonumber\\
&+\frac{8}{3}(3P_5+a^2\Delta_{IF}^{con})
\sum_Z R_{XY\pi^0\eta}^{UDS}(Z_I)l(Z_I)\Biggr],
\label{phi_PQ_disc111_su3}
\end{align}
where
\begin{align}
D_{B_1B_2\cdots B_n,B_i}^{A_1A_2\cdots A_k}(X_F)
\equiv -\frac{\partial}{\partial B_{iF}}
R_{B_1B_2\cdots B_n}^{A_1A_2\cdots A_k}(X_F)
\end{align}
%
%
and
\begin{align}
\tilde{l}(X)\equiv -\Bigl(\ln X/\Lambda^2 + 1\Bigr) + \delta_3(\sqrt{X}L).
\end{align}
The finite-volume correction $\delta_3(\sqrt{X}L)$ is~\cite{Bernard:2001yj}
\begin{align}
\delta_3(\sqrt{X}L)\equiv
2\sum_{\vec{n}\ne\vec{0}}K_0(|\vec{n}|\sqrt{X}L),
\end{align}
and $\delta_3(\sqrt{X}L)\to 0$ in infinite volume.  

A non-trivial special case of Eq.~\eqref{phi_PQ_disc111_su3} occurs for $m_x=m_y$.  We have
\begin{align}
\sigma&^{disc}_{x=y}=\frac{1}{12}\Biggl[
2\biggl(-12X_5+a^2(\Delta_{VF}^{con}
+\Delta_{VF}^{disc})\biggr)\times\nonumber\\
a^2&\delta_V^\prime\biggl(R_{X\pi^0\eta\eta^\prime}^{UDS}(X_V)\tilde{l}(X_V)+ 
\sum_Z D_{X\pi^0\eta\eta^\prime,X}^{UDS}(Z_V)l(Z_V)\biggr)\nonumber\\
-12&\,a^2\delta_V^\prime\nu_{VF}\biggl(R_{X\pi^0\eta\eta^\prime}^{UDS}(X_V)
\Bigl(l(X_V)-X_V\tilde{l}(X_V)\Bigr)\nonumber\\
&-\sum_Z Z_V D_{X\pi^0\eta\eta^\prime,X}^{UDS}(Z_V)l(Z_V)\biggr)\nonumber\\
&+(V\rightarrow A) 
+ \frac{8}{3}(3X_5+2a^2\Delta_{IF}^{con})
\biggr(R_{X\pi^0\eta}^{UDS}(X_I)\tilde{l}(X_I) \nonumber\\
&+ \sum_{Z}D_{X\pi^0\eta,X}^{UDS}(Z_I)l(Z_I)\biggr)\Biggr].
\label{pi_PQ_disc111_su3}
\end{align}
The masses $\pi^0_B$, $\eta_B$, and $\eta^\prime_B$ ($B\in\{V,A,I\}$) in
Eqs.~(\ref{phi_PQ_disc111_su3}) and (\ref{pi_PQ_disc111_su3}) are the
eigenvalues of the mass matrix in Eq.~(\ref{111_mass_matrix}).

To obtain the connected contributions in the partially quenched 2+1 flavor case, we set
$m_u=m_d$ in Eq.~\eqref{phi_full_con111_su3}.  
%
%
To obtain the disconnected contributions, we set $m_u=m_d$ in
Eq.~\eqref{disc444} and consider the two cases $m_x\neq m_y$ and $m_x=m_y$
separately.  We find
\begin{align}
\sigma&^{disc}_{x\ne y,u=d}=\frac{1}{12}\Biggl[a^4\Delta_{VF}^{con}\delta_V^\prime
\biggl(R_{X\eta\eta^\prime}^{\pi S}(X_V) \tilde{l}(X_V)\nonumber\\
&+ R_{Y\eta\eta^\prime}^{\pi S}(Y_V)\tilde{l}(Y_V) 
+ \sum_Z\Bigl(D_{X\eta\eta^\prime,X}^{\pi S}(Z_V)l(Z_V)\nonumber\\
&+ D_{Y\eta\eta^\prime,Y}^{\pi S}(Z_V)l(Z_V)\Bigr)\biggr)\nonumber\\
&+2\sum_Z\biggl(-12P_5+6Z_V\nu_{VF}
+a^2\Delta_{VF}^{disc}\biggr)\times \nonumber\\
&\quad a^2\delta_V^\prime R_{XY\eta\eta^\prime}^{\pi S}(Z_V)l(Z_V)
+(V\rightarrow A) \nonumber\\ 
&+\frac{4}{3}a^2\Delta_{IF}^{con}\biggl(R_{X\eta}^{\pi S}(X_I)\tilde{l}(X_I) 
+ R_{Y\eta}^{\pi S}(Y_I)\tilde{l}(Y_I) \nonumber\\
&+ \sum_Z\Bigl(D_{X\eta,X}^{\pi S}(Z_I)l(Z_I) 
+ D_{Y\eta,Y}^{\pi S}(Z_I)l(Z_I)\Bigr)\biggr) \nonumber\\
&+\frac{8}{3}(3P_5+a^2\Delta_{IF}^{con})
\sum_{Z}R_{XY\eta}^{\pi S}(Z_I)l(Z_I)\Biggr]
\label{K_PQ_disc2p1_su3}
\end{align}
and
%
%
\begin{align}
\sigma&^{disc}_{x=y,u=d}=\frac{1}{12}\Biggl[
2\biggl(-12X_5+a^2(\Delta_{VF}^{con}
+\Delta_{VF}^{disc})\biggr)\times\nonumber\\
a^2&\delta_V^\prime\biggl(R_{X\eta\eta^\prime}^{\pi S}(X_V)\tilde{l}(X_V)+ 
\sum_Z D_{X\eta\eta^\prime,X}^{\pi S}(Z_V)l(Z_V)\biggr)\nonumber\\
-12&\,a^2\delta_V^\prime\nu_{VF}\biggl(R_{X\eta\eta^\prime}^{\pi S}(X_V)
\Bigl(l(X_V)-X_V\tilde{l}(X_V)\Bigr)\nonumber\\
&-\sum_Z Z_V D_{X\eta\eta^\prime,X}^{\pi S}(Z_V)l(Z_V)\biggr)\nonumber\\
&+(V\rightarrow A) 
+ \frac{8}{3}(3X_5+2a^2\Delta_{IF}^{con})
\biggr(R_{X\eta}^{\pi S}(X_I)\tilde{l}(X_I) \nonumber\\
&+ \sum_Z D_{X\eta,X}^{\pi S}(Z_I)l(Z_I)\biggr)\Biggr].
\label{pi_PQ_disc2p1_su3}
\end{align}
The masses $\pi_B$, $\eta_B$, and $\eta^\prime_B$ ($B\in\{V,A,I\}$) appearing
in Eqs.~\eqref{K_PQ_disc2p1_su3} and \eqref{pi_PQ_disc2p1_su3} are the
eigenvalues of the mass matrix in Eq.~\eqref{111_mass_matrix} with $m_u=m_d$.

\subsubsection{\label{sec:Qsu3qcd}Quenched case}
The connected loop contributions are
\begin{align}
\sigma^{con}=-a^2\sum_B\,\delta_{BF}^{con}\,l(P_B)\label{phi_Q_con}.
\end{align}
%
%
To obtain the disconnected contributions, we consider Eq.~\eqref{disc444} with the replacement $D^a_{il}\to D^{a,\mathrm{quench}}_{il}$, where $D^{a,\mathrm{quench}}_{il}$ is given in Eqs.~\eqref{Dquench}-\eqref{dquench}.  We have
\begin{align}
\sigma&^{disc}_{x\ne y}=\frac{1}{12}\Biggl[a^4\Delta_{VF}^{con}\delta_V^\prime
\Bigl(\tilde{l}(X_V) + \tilde{l}(Y_V)\Bigr) \nonumber\\
&+2\Bigl(-12P_5+a^2\Delta_{VF}^{disc}\Bigr)a^2\delta_V^\prime
\frac{l(Y_V)-l(X_V)}{X_V-Y_V}\nonumber\\
&+12\,a^2\delta_V^\prime\nu_{VF}
\frac{Y_Vl(Y_V)-X_Vl(X_V)}{X_V-Y_V}
+(V\rightarrow A) \nonumber\\
&+\frac{4}{3}a^2\Delta_{IF}^{con}
\Bigl((m_0^2-\alpha X_I)\tilde{l}(X_I) + \alpha l(X_I)\nonumber\\
&+ (X\to Y)\Bigr)
+\frac{8}{3}(3P_5+a^2\Delta_{IF}^{con}) \times\nonumber\\
&\frac{(m_0^2-\alpha Y_I)l(Y_I) - (m_0^2-\alpha X_I)l(X_I)}
{X_I-Y_I}\Biggr]
\label{K_Q_disc_su3}
\end{align}
and
\begin{align}
\sigma&^{disc}_{x=y}=\frac{1}{12}\Biggl[
2\biggl(-12X_5+a^2(\Delta_{VF}^{con}
+\Delta_{VF}^{disc})\biggr)a^2\delta_V^\prime\tilde{l}(X_V)\nonumber\\
&-12\,a^2\delta_V^\prime\nu_{VF}\Bigl(l(X_V)-X_V\tilde{l}(X_V)\Bigr)\nonumber\\
&+(V\rightarrow A)
+\frac{8}{3}(3X_5+2a^2\Delta_{IF}^{con}) \times \nonumber\\
&\Bigr((m_0^2-\alpha X_I)\tilde{l}(X_I) + \alpha l(X_I)\Bigr)\Biggr],
\label{pi_Q_disc_su3}
\end{align}
where in Eq.~\eqref{K_Q_disc_su3} we substituted for the residues,
\begin{align}
R_{XY}(X_B)=-R_{XY}(Y_B)=\frac{1}{Y_B-X_B}.
\end{align}
The loop contributions to the pion and kaon masses in the case of three non-degenerate quarks and in the isospin limit can be obtained from Eqs.~\eqref{phi_Q_con}, \eqref{K_Q_disc_su3}, and \eqref{pi_Q_disc_su3} by appropriately choosing $m_x$ and $m_y$.
%
%

\subsection{$SU(2)$ chiral perturbation theory}
Expansions about the $SU(2)$ chiral limit are often better behaved than expansions about the $SU(3)$ chiral limit.  The corresponding $\chi$PT was developed in Ref.~\cite{Gasser:1983yg} and extended in Refs.~\cite{Du:2009ih,Bazavov:2010yq} to describe results obtained with rooted staggered fermions.  To date $SU(2)$ S$\chi$PT analyses have been restricted to the taste Goldstone sector~\cite{Bazavov:2009fk,Bazavov:2009ir,Bazavov:2010yq}.  Beginning with the loop contributions from $SU(3)$ S$\chi$PT, we write down corresponding $SU(2)$ S$\chi$PT loop contributions to the taste non-Goldstone PGB masses.

\subsubsection{\label{sec:fullsu2qcd}Fully dynamical case}
From Eqs.~\eqref{piplus_full_con111_su3}, \eqref{cK_full_con111_su3}, and \eqref{nK_full_con111_su3}, we have
\begin{align}
\sigma^{con}_{\pi^+} = -a^2\sum_B\Biggl(\delta_{BF}^{con}\,l(\pi^+_B)\,+&\,
\frac{\Delta_{BF}^{con}}{48}\Bigl(l(U_B) + 2l(\pi^+_B) \nonumber\\
& + l(D_B)\Bigr) \Biggr)\label{piplus_full_con111_su2}\\
\sigma^{con}_{K^+} = -a^2\sum_B
\frac{\Delta_{BF}^{con}}{48}\Bigl(l(U_B&)
+ l(\pi^+_B)\Bigr)\label{cK_full_con111_su2}\\
\sigma^{con}_{K^0} = -a^2\sum_B
\frac{\Delta_{BF}^{con}}{48}\Bigl(l(\pi^+_B&)
+ l(D_B)\Bigr).\label{nK_full_con111_su2}
\end{align}

To obtain the disconnected loop contributions in the 1+1+1 flavor case, we consider Eqs.~\eqref{pi_full_disc111_su3}, \eqref{cK_full_disc111_su3}, \eqref{nK_full_disc111_su3}, and Eq.~\eqref{disc444}.  We find
\begin{align}
\sigma&^{disc}_{\pi^+}=\frac{1}{12}\Biggl[a^4\Delta_{VF}^{con}\,\delta_V^\prime
\sum_{X}\biggl(R_{U\pi^0\eta}^{D}(X_V)l(X_V) \nonumber\\
&+ R_{D\pi^0\eta}^{U}(X_V)l(X_V)\biggr)
\nonumber\\
&+2\sum_{X}\biggl(-12\pi_5^{+}+6X_V\nu_{VF}+a^2\Delta_{VF}^{disc}\biggr)\times \nonumber\\
&\mbox{\quad}a^2\delta_V^\prime R_{\pi^0\eta}(X_V)l(X_V)+(V\rightarrow A) \nonumber\\
&+2a^2\Delta_{IF}^{con}\sum_{X}\biggl(R_{U\pi^0}^{D}(X_I)l(X_I) + R_{D\pi^0}^{U}(X_I)l(X_I)\biggr) \nonumber\\
&+4(3\pi_5^{+}+a^2\Delta_{IF}^{con})l(\pi^0_I)\Biggr],
\label{pi_full_disc111_su2}
\end{align}
\begin{align}
\sigma^{disc}_{K^+}&=\frac{1}{12}\Biggl[a^4\Delta_{VF}^{con}\delta_V^\prime
\sum_{X}\,R_{U\pi^0\eta}^{D}(X_V)l(X_V)\nonumber\\
&-12\sum_{X}\,a^2\delta_V^\prime R_{\pi^0\eta}^{D}(X_V)l(X_V)+(V\rightarrow A) \nonumber\\
&+2a^2\Delta_{IF}^{con}\sum_{X}\,R_{U\pi^0}^{D}(X_I)l(X_I)\nonumber\\
&+6(D_I-\pi_I^0)l(\pi^0_I)\Biggr],
\label{cK_full_disc111_su2}
\end{align}
\begin{align}
\sigma^{disc}_{K^0}&=\frac{1}{12}\Biggl[a^4\Delta_{VF}^{con}\delta_V^\prime \sum_{X}\,R_{D\pi^0\eta}^{U}(X_V)l(X_V)\nonumber\\
&-12\sum_{X}\,a^2\delta_V^\prime R_{\pi^0\eta}^{U}(X_V)l(X_V)+(V\rightarrow A) \nonumber\\
&+2a^2\Delta_{IF}^{con}\sum_{X}\,R_{D\pi^0}^{U}(X_I)l(X_I) \nonumber\\
&+6(U_I-\pi^0_I)l(\pi^0_I)\Biggr].
\label{nK_full_disc111_su2}
\end{align}
The taste vector and axial residues in Eqs.~\eqref{pi_full_disc111_su2}-\eqref{nK_full_disc111_su2} can be simplified using the tree-level masses:
\begin{align}
\pi^0_B=&\frac{1}{2}(U_B+D_B)+\frac{a^2\delta_B^\prime}{4}\nonumber\\
&-(\mr{sgn}\,\delta_B^\prime)\frac{1}{2}\sqrt{(D_B-U_B)^2+\frac{1}{4}(a^2\delta_B^\prime)^2}\\
\eta_B=&\frac{1}{2}(U_B+D_B)+\frac{a^2\delta_B^\prime}{4}\nonumber\\
&+(\mr{sgn}\,\delta_B^\prime)\frac{1}{2}\sqrt{(D_B-U_B)^2+\frac{1}{4}(a^2\delta_B^\prime)^2}
\end{align}
for $B=V,A$.
We have
\begin{align}
R_{U\pi^0\eta}^{D}(U_B)&=\frac{4}{a^2\delta_B^\prime}\\
R_{U\pi^0\eta}^{D}(\pi^0_B)&=-\frac{2}{a^2\delta_B^\prime}(1+\sin\beta_B)\\
R_{U\pi^0\eta}^{D}(\eta_B)&=-\frac{2}{a^2\delta_B^\prime}(1-\sin\beta_B)\\
R_{D\pi^0\eta}^{U}(D_B)&=\frac{4}{a^2\delta_B^\prime}\\
R_{D\pi^0\eta}^{U}(\pi^0_B)&=-\frac{2}{a^2\delta_B^\prime}(1-\sin\beta_B)\\
R_{D\pi^0\eta}^{U}(\eta_B)&=-\frac{2}{a^2\delta_B^\prime}(1+\sin\beta_B)\\
R_{\pi^0\eta}(\pi^0_B)&=\frac{2}{a^2\delta_B^\prime}\cos\beta_B\\
R_{\pi^0\eta}(\eta_B)&=-\frac{2}{a^2\delta_B^\prime}\cos\beta_B\\
R_{\pi^0\eta}^{D}(\pi^0_B)&=\frac{1}{2}\left(1+\sin\beta_B-\cos\beta_B\right)\\
R_{\pi^0\eta}^{D}(\eta_B)&=\frac{1}{2}\left(1-\sin\beta_B+\cos\beta_B\right),
\end{align}
where 
\begin{align}
\sin\beta_B&\equiv(\mr{sgn}\,\delta_B^\prime)\frac{D_B-U_B}{\sqrt{(D_B-U_B)^2+\frac{1}{4}(a^2\delta_B^\prime)^2}}\\
\cos\beta_B&\equiv(\mr{sgn}\,\delta_B^\prime)\frac{\frac{1}{2}a^2\delta_B^\prime}{\sqrt{(D_B-U_B)^2+\frac{1}{4}(a^2\delta_B^\prime)^2}}.
\end{align}
In the isospin limit, $\beta_B=0$.

The connected loops in the 2+1 flavor case are
%
%
\begin{align}
\sigma^{con}_{\pi} &= -a^2\sum_B\left(\delta_{BF}^{con} + \frac{\Delta_{BF}^{con}}{12}\right)l(\pi_B)\label{pi_full_con2p1_su2}\\
\sigma^{con}_{K} &= -a^2\sum_B
\frac{\Delta_{BF}^{con}}{24}l(\pi_B),\label{K_full_con2p1_su2}
\end{align}
%
%
%
and the disconnected loops are
\begin{align}
\sigma^{disc}_{\pi}=\frac{1}{12}&\biggl[4\Bigl(-12\pi_5 + a^2(\Delta_{VF}^{con} + \Delta_{VF}^{disc})\Bigr)\times\nonumber\\
&\phantom{+}(l(\pi_V)-l(\eta_V))\nonumber\\
& + 24\nu_{VF}(\pi_V l(\pi_V)-\eta_V l(\eta_V))\nonumber\\
& + (V\rightarrow A) \nonumber\\
& +4(3\pi_5+2a^2\Delta_{IF}^{con})l(\pi_I)\biggr],
\label{isopi_full_disc2p1_su2}
\end{align}
\begin{align}
\sigma^{disc}_{K}=\frac{1}{12}&\biggl[2a^2\Delta_{VF}^{con}(l(\pi_V)-l(\eta_V))
-12\,a^2\delta_V^\prime l(\eta_V)\nonumber\\
&+(V\rightarrow A)+2a^2\Delta_{IF}^{con}l(\pi_I)\biggr].
\label{K_full_disc2p1_su2}
\end{align}
The masses in Eqs.~\eqref{pi_full_con2p1_su2}, \eqref{K_full_con2p1_su2}, \eqref{isopi_full_disc2p1_su2}, and \eqref{K_full_disc2p1_su2} are
\begin{align}
\pi_B&=2\mu m_u + a^2\Delta_B\quad \forall\ B\\
\eta_B&=2\mu m_u + a^2\Delta_B + \frac{a^2\delta_B^\prime}{2}\quad B\in\{V,A\}.
\end{align}
All mesons circulating in loops in the $SU(2)$ chiral theory are pions.

\subsubsection{\label{sec:PQsu2qcd}Partially quenched case}
We obtain the connected contributions in the 1+1+1 flavor case by dropping
terms in Eq.~\eqref{phi_full_con111_su3} corresponding to loops with a strange
sea quark; {\it i.e.}, the sum over $Q$ excludes the $xs$ and $ys$ mesons, and
we treat the $x$ and $y$ quarks as light.\footnote{Another 
case of interest would be that of a single heavy (strange) valence quark, $m_y\sim m_s\gg m_x\sim m_{u,d}$.}
%
%
To obtain the disconnected contributions in the 1+1+1 flavor case, we consider Eqs.~\eqref{phi_PQ_disc111_su3}, \eqref{pi_PQ_disc111_su3}, and \eqref{disc444}.  We have
\begin{align}
\sigma&^{disc}_{x\neq y}=\frac{1}{12}\Biggl[a^4\Delta_{VF}^{con}\delta_V^\prime
\biggr(R_{X\pi^0\eta}^{UD}(X_V) \tilde{l}(X_V)\nonumber\\
&+ R_{Y\pi^0\eta}^{UD}(Y_V)\tilde{l}(Y_V) + \sum_Z\Bigl(D_{X\pi^0\eta,X}^{UD}(Z_V)l(Z_V)\nonumber\\
&+ D_{Y\pi^0\eta,Y}^{UD}(Z_V)l(Z_V)\Bigr)\biggl)\nonumber\\
&+2\sum_Z\biggl(-12P_5+6Z_V\nu_{VF}+a^2\Delta_{VF}^{disc}\biggr)\times \nonumber\\
&\quad a^2\delta_V^\prime R_{XY\pi^0\eta}^{UD}(Z_V)l(Z_V)+(V\rightarrow A) \nonumber\\
&+2a^2\Delta_{IF}^{con}
\biggl(R_{X\pi^0}^{UD}(X_I)\tilde{l}(X_I)
+ R_{Y\pi^0}^{UD}(Y_I)\tilde{l}(Y_I) \nonumber\\
&+ \sum_Z\Bigl(D_{X\pi^0,X}^{UD}(Z_I)l(Z_I)
+ D_{Y\pi^0,Y}^{UD}(Z_I)l(Z_I)\Bigr)\biggr) \nonumber\\
&+4(3P_5+a^2\Delta_{IF}^{con})
\sum_Z R_{XY\pi^0}^{UD}(Z_I)l(Z_I)\Biggr]
\label{phi_PQ_disc111_su2}
\end{align}
and
%
%
\begin{align}
\sigma&^{disc}_{x=y}=\frac{1}{12}\Biggl[
2\biggl(-12X_5+a^2(\Delta_{VF}^{con}
+\Delta_{VF}^{disc})\biggr)\times\nonumber\\
&a^2\delta_V^\prime\biggl(R_{X\pi^0\eta}^{UD}(X_V)\tilde{l}(X_V)+
\sum_Z D_{X\pi^0\eta,X}^{UD}(Z_V)l(Z_V)\biggr)\nonumber\\
&-12\,a^2\delta_V^\prime \nu_{VF}\biggl(R_{X\pi^0\eta}^{UD}(X_V)
\Bigl(l(X_V)-X_V\tilde{l}(X_V)\Bigr)\nonumber\\
&-\sum_Z Z_V D_{X\pi^0\eta,X}^{UD}(Z_V)l(Z_V)\biggr)\nonumber\\
&+(V\rightarrow A)
+ 4(3X_5+2a^2\Delta_{IF}^{con})
\biggl(R_{X\pi^0}^{UD}(X_I)\tilde{l}(X_I) \nonumber\\
&+ \sum_Z D_{X\pi^0,X}^{UD}(Z_I)l(Z_I)\biggr)\Biggr].
\label{pi_PQ_disc111_su2}
\end{align}
Setting $m_u=m_d$, we have the disconnected contributions in the 2+1 flavor case:
\begin{align}
\sigma&^{disc}_{x\neq y,u=d}=\frac{1}{12}\Biggl[a^4\Delta_{VF}^{con}\delta_V^\prime
\biggr(R_{X\eta}^{\pi}(X_V) \tilde{l}(X_V)\nonumber\\
&+ R_{Y\eta}^{\pi}(Y_V)\tilde{l}(Y_V) + \sum_Z\Bigl(D_{X\eta,X}^{\pi}(Z_V)l(Z_V)\nonumber\\
&+ D_{Y\eta,Y}^{\pi}(Z_V)l(Z_V)\Bigr)\biggl)\nonumber\\
&+2\sum_Z\biggl(-12P_5+6Z_V\nu_{VF}+a^2\Delta_{VF}^{disc}\biggr)\times \nonumber\\
&\quad a^2\delta_V^\prime R_{XY\eta}^{\pi}(Z_V)l(Z_V)+(V\rightarrow A) \nonumber\\
&+2a^2\Delta_{IF}^{con}
\biggl(\Bigl(l(X_I)+(\pi_I-X_I)\tilde{l}(X_I)\Bigr)\nonumber\\
&+\Bigl(l(Y_I)+(\pi_I-Y_I)\tilde{l}(Y_I)\Bigr)\biggr)\nonumber\\
&+4(3P_5+a^2\Delta_{IF}^{con})
\sum_Z R_{XY}^{\pi}(Z_I)l(Z_I)\Biggr]
\label{K_PQ_disc2p1_su2}
\end{align}
and
%
%
%
\begin{align}
\sigma&^{disc}_{x=y,u=d}=\frac{1}{12}\Biggl[
2\biggl(-12X_5+a^2(\Delta_{VF}^{con}
+\Delta_{VF}^{disc})\biggr)\times\nonumber\\
&a^2\delta_V^\prime\biggl(R_{X\eta}^{\pi}(X_V)\tilde{l}(X_V)+
\sum_Z D_{X\eta,X}^{\pi}(Z_V)l(Z_V)\biggr)\nonumber\\
&-12\,a^2\delta_V^\prime \nu_{VF}\biggl(R_{X\eta}^{\pi}(X_V)
\Bigl(l(X_V)-X_V\tilde{l}(X_V)\Bigr)\nonumber\\
&-\sum_Z Z_V D_{X\eta,X}^{\pi}(Z_V)l(Z_V)\biggr)
+(V\rightarrow A)\nonumber\\
&+ 4(3X_5+2a^2\Delta_{IF}^{con})
\Bigl(l(X_I)+(\pi_I-X_I)\tilde{l}(X_I)\Bigr)\Biggr].
\label{pi_PQ_disc2p1_su2}
\end{align}
%
%

\section{\label{sec:disc}Conclusion}
Our final results for the masses of the flavor-charged PGBs through NLO in
S$\chi$PT are given by adding Eq.~(\ref{SelfEDiv}) evaluated on-shell to the
tree-level (LO) result of Eq.~(\ref{eq:tm}).  These results and others of
interest can be obtained from those in the 4+4+4 flavor theory given in
Eqs.~\eqref{fin_con444}, \eqref{disc444}, and \eqref{eq:sanal444} of
Sec.~\eqref{sec:res444}.  Applying the replica method to reduce the number of
tastes per flavor from four to one gives the connected tadpole, disconnected
tadpole, and NLO (analytic) tree-level contributions to the on-shell self-energies.
In Sec.~\ref{sec:res} we write down the connected and disconnected tadpoles
in the 1+1+1 flavor and 2+1 flavor cases in $SU(3)$ and $SU(2)$ $\chi$PT.

For the fully dynamical case with three non-degenerate quarks, the results in
the $SU(3)$ chiral theory are in Eqs.~\eqref{piplus_full_con111_su3} through
\eqref{nK_full_disc111_su3}.  The corresponding results in the isospin limit
are in Eqs.~\eqref{pi_full_con2p1_su3} through \eqref{K_full_disc2p1_su3}.  
Expansions about the $SU(2)$ chiral limit are given in
Eqs.~\eqref{piplus_full_con111_su2} through \eqref{nK_full_disc111_su2} and
Eqs.~\eqref{pi_full_con2p1_su2} through \eqref{K_full_disc2p1_su2}.
For the quenched case, the results are in Eqs.~\eqref{phi_Q_con} through
\eqref{pi_Q_disc_su3}, where the LECs are the quenched counterparts of those
in the theories with dynamical quarks.

For the partially quenched case, the connected contributions have the same form
as those in the fully dynamical case, Eq.~\eqref{phi_full_con111_su3}.  For
three non-degenerate sea quarks, the disconnected contributions in the $SU(3)$
chiral theory are in Eqs.~\eqref{phi_PQ_disc111_su3} and
\eqref{pi_PQ_disc111_su3}.  Taking the isospin limit in the sea, the
corresponding results are in Eqs.~\eqref{K_PQ_disc2p1_su3} and
\eqref{pi_PQ_disc2p1_su3}.  The expansions about the $SU(2)$ chiral limit are
given in Eqs.~\eqref{phi_PQ_disc111_su2}-\eqref{pi_PQ_disc111_su2} and
Eqs.~\eqref{K_PQ_disc2p1_su2}-\eqref{pi_PQ_disc2p1_su2}.

The LO contributions to the masses break taste $SU(4)_T$ to taste
$SO(4)_T$~\cite{Lee:1999zxa}.  At NLO the (tadpole) loops respect taste
$SO(4)_T$, the tree-level counterterms from the Gasser-Leutwyler Lagrangian
respect taste $SU(4)_T$, and tree-level counterterms from the Sharpe-Van de
Water Lagrangian break spacetime-taste $SO(4)\times SU(4)_T$ to the lattice
symmetry, $\Gamma_4\rtimes SW_{4,\mathrm{diag}}$.

\begin{figure}[tbhp]
  \includegraphics[width=20pc]{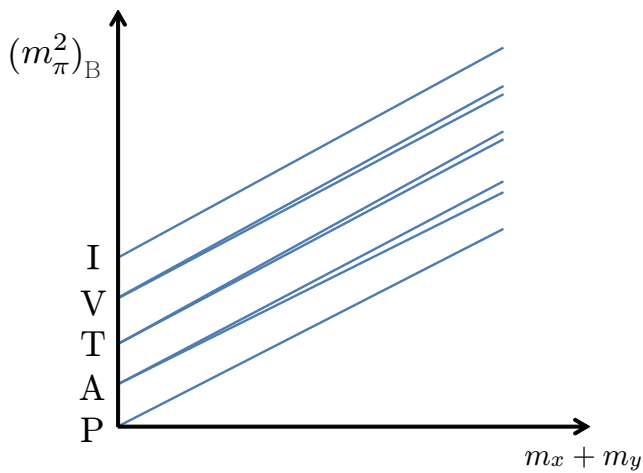}
  \caption{Pattern of taste symmetry breaking in squared PGB masses.  The PGBs fall into 8 lattice irreps, and the masses receive corrections to the slopes and intercepts that lift the degeneracy of the taste $SO(4)_T$ irreps.  The corrections to the intercepts are of $\mc{O}(a^4)$ and very small on typical lattices.
  }
  \label{fig:m_pi_sq}
\end{figure}

The pattern of taste symmetry breaking is illustrated in
Fig.~\ref{fig:m_pi_sq}.  Regarded as functions of the valence masses, the LO masses receive corrections at NLO to their slopes and intercepts.  The chiral logarithms contribute to both types of corrections but do not lift degeneracies within taste $SO(4)_T$ irreps.  A small subset of the Sharpe-Van de Water counterterms breaks the $SO(4)_T$ symmetry.  With HYP-smeared staggered valence quarks on MILC coarse lattices, the corrections to the intercepts are smaller than the statistical uncertainties~\cite{Bae:2008qe,Adams:2008ta}; Fig.~\ref{fig:m_pi_sq} represents this case.  The exact chiral symmetry at nonzero lattice spacing ensures corrections to the intercept of the taste Goldstone ($P$) mesons vanish.

As discussed in Appendix~\ref{app:pattern}, the $SO(4)_T$-breaking contributions to the masses of the flavor-charged PGBs arise from only three operators in the Sharpe-Van de Water Lagrangian~\cite{Sharpe:2004is}.  As a direct consequence, the $SO(4)_T$-breaking corrections to the slopes and intercepts of Fig.~\ref{fig:m_pi_sq} depend on only three LECs.  One could obtain them from the splittings in the slopes of the taste vector, axial, and tensor irreps.  

The LECs of the LO taste-breaking potentials enter the results in four specific
ways: The couplings of the potential $\mathcal{U}$, Eq.~\eqref{TBPot1}, enter
only {\it via} the tree-level mass splittings and the coefficients of the
chiral logarithms from disconnected tadpoles; for each taste channel, these
coefficients are given in Tables~\ref{ConDiag} and \ref{DiscDiag}.  The
couplings of the potential $\mathcal{U}^\prime$, Eq.~\eqref{TBPot2}, enter only
{\it via} the hairpin coefficients of the Goldstone sector and the two linear
combinations of Eq.~\eqref{eq:plus_hairpin}.  The former arise in disconnected
propagators, while the latter multiply connected tadpoles with valence-valence mesons
in the loop; for each taste channel, they are given in Table~\ref{ConDiag2}.

From Eqs.~\eqref{Itsplitdef}-\eqref{Ptsplitdef} and Tables~\ref{ConDiag} and
\ref{DiscDiag}, we observe that the coefficients of the chiral logarithms from
the disconnected tadpoles are completely determined by the tree-level mass
splittings.  The tree-level mass splittings also determine the coefficients of
chiral logarithms from connected tadpoles with sea quarks in the loop.  Having
determined the $SO(4)_T$-breaking terms and the LO masses, one could perform
fits to partially quenched data to extract the remaining (two) coefficients of
the connected contributions and the coefficients of the $SO(4)_T$-preserving
analytic corrections at NLO.  We note in passing that the coefficients of the
quenched and $SU(2)$ chiral theories are different from the coefficients of the
$SU(3)$ chiral theory.

The calculation here can be extended to mixed action $\chi$PT and to
other quantities of phenomenological interest such as decay constants,
form factors, and mixing parameters.  For example, one could consider
HISQ or HYP-smeared staggered on asqtad staggered simulations; the
S$\chi$PT for both cases is the
same~\cite{Bar:2003mh,Bar:2005tu,Chen:2007ug,Chen:2009su}.  A
calculation for $B_K$ is given in Ref.~\cite{Bae:2010ki}.  We plan to
calculate in the near future the one-loop corrections to the mass
spectrum of pions and kaons in the mixed action case.

\begin{acknowledgments}
  This research was supported by the Creative Research Initiatives program
(3348-20090015) of the NRF grant of the Korean government (MEST).  We thank
Steve Sharpe for hospitality during our visit at the University of Washington.
We thank Steve Sharpe and Claude Bernard for helpful discussions.
\end{acknowledgments}

\appendix
\section{\label{app:pcf}Power counting formula}
In Sec.~\ref{sec:pow} we recalled the standard power counting of S$\chi$PT.
Here we derive the power counting formula, Eq.~\eqref{eq:power}.  The
derivation is based closely on the discussion for the continuum case in
Ref.~\cite{Scherer:2002tk}.

We begin by noting that the effective continuum Symanzik action contains no
operators of mass dimension five or seven~\cite{Sharpe:2004is}:  
\begin{align}
S_\mr{SYM} = S_4 + a^2 S_6 + a^4 S_8 + \dots
\end{align} 
For the following discussion, we assume without proof that no operators of odd
mass dimension appear at higher orders in the Symanzik action.  At the end of
the derivation, we consider the restrictions this assumption places on the
validity of Eq.~\eqref{eq:power}.

Mapping the operators of the Symanzik action into the Lagrangian of S$\chi$PT
and using it to compute an arbitrary amplitude, we note that dependence on the
lattice spacing enters {\it via} the vertices and the (tree-level) propagators.
By the assumption of the previous paragraph, all vertices and propagators
depend analytically on $a^2$.  The Symanzik and S$\chi$PT actions are
translation invariant, so momentum conservation holds in S$\chi$PT, as in the continuum theory.

The lattice-spacing dependence of the propagators can be deduced from
Eqs.~\eqref{prop}, \eqref{D_piece}, \eqref{hpV_hpA}, \eqref{eq:fntm},
\eqref{eq:tm}, \eqref{444_mass_matrix}, and Eqs.~\eqref{hpdef} through
\eqref{Ptsplitdef}.  The propagators receive corrections proportional to the
hairpin couplings of Eqs.~\eqref{hpdef} and \eqref{hpdefA} and the taste
splittings of Eqs.~\eqref{Itsplitdef} through \eqref{Ptsplitdef}.  The former
enter only the disconnected parts of the flavor-neutral propagators; the latter
are corrections to the tree-level masses of all the PGBs.

In any given amplitude, internal lines contribute factors of
\begin{align}
&\int \frac{d^4q}{(2\pi)^4}
\langle\phi^a_{ij}\phi^b_{kl}\rangle\\
=\delta^{ab}&\int \frac{d^4q}{(2\pi)^4}\left(\delta_{il}\delta_{jk}\frac{1}{q^2+\frac{1}{2}(I_a+J_a)}+\delta_{ij}\delta_{kl}D^a_{il}\right),
\end{align}
and vertices contribute momentum-conserving delta functions and couplings of the form
\begin{align}
v=p^{2n_{p^2}-n_q}\,q^{n_q}\,m_q^{n_m}\,a^{2n_{a^2}},\label{eq:vrt_frm}
\end{align}
where $2n_{p^2}$ is the number of derivatives in the interaction, $p$ is an
external momentum, $q$ is an internal momentum, $n_q$ is the number of internal
lines contracted with the vertex, $n_m$ is the number of quark-mass factors
from the vertex, $m_q$ is a light ($u$, $d$, or $s$) quark mass, $2n_{a^2}$ is
the number of lattice-spacing factors from the vertex, and $a$ is the lattice
spacing.  

Rescaling the external momenta, quark masses, and lattice spacing,
\begin{align}
p&\to \sqrt{\varepsilon}\label{eq:p_scale}\\
m_q&\to \varepsilon m_q\\
a^2&\to \varepsilon a^2,
\end{align}
the factors from internal lines scale as
\begin{align}
\int \frac{d^4q}{(2\pi)^4}\langle\phi^a_{ij}\phi^b_{kl}\rangle\to
\varepsilon\int\frac{d^4q}{(2\pi)^4}\langle\phi^a_{ij}\phi^b_{kl}\rangle,
\end{align}
where we changed the variable of integration from $q$ to $\sqrt{\varepsilon}q$.
Under the same change of variable and the rescaling of Eq.~\eqref{eq:p_scale}, the momentum-conserving delta functions
scale as
\begin{align}
\delta^4(p+q)&\to\delta^4(\sqrt{\varepsilon}(p+q))\\
&=\frac{1}{\varepsilon^2}\delta^4(p+q),
\end{align}
while the vertex contribution of Eq.~\eqref{eq:vrt_frm} scales as
\begin{align}
v&\to
\varepsilon^{n_{p^2}+n_m+n_{a^2}}v\\
&=\varepsilon^nv,\label{eq:v_scale}
\end{align}
where $n\equiv n_{p^2}+n_m+n_{a^2}$.  Recalling the definitions of $n_{p^2}$,
$n_m$, and $n_{a^2}$ and the organization of the Lagrangian in
Eq.~\eqref{LagExp}, we conclude that Eq.~\eqref{eq:v_scale} implies that a
vertex contribution scales with a factor of $\varepsilon^n$ if and only if the
vertex is from an interaction in the Lagrangian $\mc{L}_{2n}$ of
Eq.~\eqref{LagExp}.

A given Feynman graph $\mc{M}(p_i,m_q,a^2)$ has $N_I$ internal lines and $N_V$ vertices, where
\begin{align}
N_V=\sum_{n=1}^\infty N_{2n};
\end{align}
{\it i.e.}, the number of vertices in the graph is the sum of the number of
vertices $N_{2n}$ from each term $\mc{L}_{2n}$ in the Lagrangian.  Because the total
momentum flowing into the diagram equals the total momentum flowing out, one
momentum-conserving delta function does not contribute an independent
constraint and is factored out of the amplitude.  Multiplying the rescaling
factors of the internal lines, vertex contributions, and remaining $N_V-1$
momentum-conserving delta functions gives
\begin{align}
\mc{M}(p_i,m_q,a^2)\to\varepsilon^D\mc{M}(p_i,m_q,a^2)
\end{align}
where
\begin{align}
D=N_I-2(N_V-1)+\sum_{n=1}^\infty nN_{2n}.\label{eq:almost_D}
\end{align}
The number of loops in a diagram is the number of independent integrations
after imposing the $N_V-1$ constraints from the momentum-conserving delta
functions:
\begin{align}
N_L=N_I-(N_V-1).
\end{align}
Using this relation to eliminate $N_I$ from Eq.~\eqref{eq:almost_D} gives the
desired result:
\begin{align}
D&=N_L-N_V+1+\sum_{n=1}^\infty nN_{2n}\\
&=1+N_L+\sum_{n=1}^\infty (n-1)N_{2n}.
\end{align}

We now reflect on the validity of this result.  We assumed that no operators of
odd mass dimension appear at any order in the Symanzik action.  This assumption
is known to be true only through mass dimension eight.  In principle it could
be violated at mass dimension nine:
\begin{align}
S_\mr{SYM}=S_4 + a^2 S_6 + a^4 S_8 + a^5 S_9 + \dots
\end{align}

In this case the rescaling is the same for the internal lines and the
momentum-conserving delta functions, but the possible vertex contributions are
different:
\begin{align}
v=p^{n_p-n_q}\,q^{n_q}\,m_q^{n_m}\,a^{n_a},
\end{align}
and the expansion of the S$\chi$PT Lagrangian can be written
\begin{align}
\mc{L}=\sum_{n=2,4,5,\dots}\mc{L}_n.
\end{align}
We allow for the number of derivatives $n_p$ in an operator to be odd because
they are the only objects in the chiral Lagrangian with indices that can
contract with those of taste matrices to construct operators with an odd number
of taste spurions.

The vertex factors now rescale as
\begin{align}
v\to\varepsilon^{n/2}v,
\end{align}
where $n\equiv n_p+2n_m + n_a$, and we have
\begin{align}
N_V=\sum_n N_n.
\end{align}
The modified power counting formula is
\begin{align}
D = 1 + N_L + \sum_{n=2,4,5,\dots}\left(\frac{n}{2}-1\right)N_n,
\end{align}
and writing out the solutions to this equation for $D=1,2,\tfrac{5}{2},3$
yields the same solutions to the power counting relation for $D=1,2,3$ as
before.  For $D=1,2$, the new operators in $\mc{L}_5$ do not contribute, and
the power counting of Eq.~\eqref{eq:power} is justified through NLO.  
%
%

\section{\label{app:pattern}NLO analytic corrections and taste symmetry breaking}
The NLO analytic corrections to the PGB masses break spacetime-taste $SO(4)\times SO(4)_T$ to the diagonal hypercubic subgroup $SW_\mathrm{4,diag}$ of the lattice theory.  Here we consider the responsible operators and note the pattern of symmetry breaking in the mass spectrum.

Sharpe and Van de Water enumerated the NLO Lagrangian giving rise to the NLO analytic corrections~\cite{Sharpe:2004is}.  Although many operators contribute at NLO, the vast majority respect the remnant taste symmetry $SO(4)_T$; only three operators are responsible for the symmetry breaking in the masses of the flavor-charged PGBs, and all are of type $(n_{p^2},n_m,n_{a^2})=(1,0,1)$.  For example, the operator
\begin{align}
\frac{a^2C_{36V}}{f^2}\sum_\mu\,\Tr(\partial_\mu\phi\xi_\mu\partial_\mu\phi\xi_\mu)
\end{align}
yields the correction to the self-energy for $\phi_{xy}^t$
\begin{align}
\frac{8a^2C_{36V}}{f^2}\sum_\mu\,p_\mu p_\mu \theta^{\mu t},
\end{align}
which in the dispersion relations breaks (spatial) rotation invariance and lifts the $SO(4)_T$ degeneracies of the masses of the PGBs.  

The symmetry breaking corrections to the dispersion relations were
calculated by Sharpe and Van de Water.  They have the form
\begin{align}
E_I^2&=\vec{p}^2
+M_{I}^2(1+\kappa_{I})\\
E_4^2&=\vec{p}^2(1+\kappa_4-\kappa_i)
+M_{V}^2(1+\kappa_4)\\
E_i^2&=p_i^2(1+\kappa_i-\kappa_4)+p_j^2+p_k^2
+M_{V}^2(1+\kappa_i)\\
E_{ij}^2&=(p_i^2+p_j^2)(1+\kappa_{ij}-\kappa_{i4})+p_k^2
+M_{T}^2(1+\kappa_{ij})\\
E_{i4}^2&=p_i^2+(p_j^2+p_k^2)(1+\kappa_{i4}-\kappa_{ij})
+M_{T}^2(1+\kappa_{i4})\\
E_{i5}^2&=p_i^2(1+\kappa_{i5}-\kappa_{45})+p_j^2+p_k^2
+M_{A}^2(1+\kappa_{i5})\\
E_{45}^2&=\vec{p}^2(1+\kappa_{45}-\kappa_{i5})
+M_{A}^2(1+\kappa_{45})\\
E_5^2&=\vec{p}^2 + M_{P}^2(1+\kappa_5),
\end{align}
where $M_t$ is the flavor-charged PGB mass through NLO including all but the
taste $SO(4)_T$ violating contributions, and we use $\kappa_t$ to denote the
$\delta_t$ of Ref.~\cite{Sharpe:2004is}.  Writing out the various contributions
to $M_t$ more explicitly,
\begin{align}
M_t^2=m_{xy}^2+a^2\Delta_t+\ell_t + a_t,
\end{align}
where the first two terms are the LO result, $\ell_t$ is the sum of
all loop corrections for taste $t$, and $a_t$ is the sum of all NLO
analytic corrections respecting taste $SO(4)_T$.  The results of
Sharpe and Van de Water then imply that the NLO $SO(4)_T$-breaking
corrections are
\begin{align}
(m_{xy}^2+a^2\Delta_t)\kappa_t=m_{xy}^2\kappa_t+a^2\Delta_t\kappa_t.
\end{align}
Since the corrections $\kappa_t$ are proportional to $a^2$, the NLO
masses of the flavor-charged PGBs, considered as functions of
$m_x+m_y$, receive corrections to their slopes and intercepts.
However, these corrections are completely determined by only three
{\it a priori} unknown LECs in the Sharpe-Van de Water Lagrangian.

Recalling the $SO(4)_T$-breaking corrections in Eq.~\eqref{eq:sanal444},
\begin{align}
\Xi_t = \frac{16a^2}{f^2}(m_{xy}^2\mc{E}_t+a^2\mc{G}_t),\label{eq:EG}
\end{align}
we see that $\Xi_t$ contains the 16 {\it a priori} independent
coefficients $\mc{E}_t$ and $\mc{G}_t$, and that the part of $\Xi_t$ which breaks
$SO(4)_T$ is fixed by only three LECs in the Sharpe-Van de Water
Lagrangian.  The remaining part, which is $SO(4)_T$-symmetric, can be
determined by fitting as discussed in Sec.~\ref{sec:disc}.

In summary, the $SO(4)_T$-breaking corrections to the dispersion
relations imply the presence of $\mathcal{O}(a^2m_q)$ and
$\mathcal{O}(a^4)$ analytic corrections to the masses of the
corresponding (flavor-charged) PGBs.  These corrections are determined
by only three LECs, which may be taken to be the splittings of the
$SO(4)_T$ vector, tensor, and axial irreps.

\section{\label{app:eg}Example loop calculations}
In Sec.~\ref{sec:vres} we wrote down the results for each class of vertices in
the graphs contributing to the PGB self-energies at NLO.  Here we detail the
calculation of these results in two cases:  for the (mass) vertices of
Eq.~(\ref{mass4pt}), which yield Eq.~(\ref{mass4ptloops}), and the
($a^2\mathcal{U}$) vertices of Eqs.~(\ref{C1_4pt}) through (\ref{C4_4pt}),
which yield Eq.~(\ref{C4ptloops}) with Eqs.~(\ref{dat_def}) and
(\ref{eat_def}).

Perhaps the approach taken here can be extended without too much difficulty
to calculations beyond NLO.  The integrals associated with two-loop
contributions will differ from those entering tadpole graphs, but in principle
one can construct them from the propagator (Eq.~(\ref{prop})) and the generic
forms of loops with (for example) $\CO(\phi^6)$ vertex classes from the LO
Lagrangian.
\subsection{Mass vertices}
We begin by considering the vertices in Eq.~(\ref{mass4pt}):
\begin{equation}
-\frac{\mu}{48f^2}\,\tau_{abcd}\,m_i\,\phi^a_{ij}\phi^b_{jk}\phi^c_{kl}\phi^d_{li}\label{m4pt_rem}
\end{equation}
The corresponding tadpole graphs are the sum of all complete contractions with
the external fields $\phi_{xy}^t$ and $\phi_{yx}^t$.  Contractions between a
field in the vertex and an external field vanish unless the flavor and taste
indices match, in which case they contribute factors of unity to amputated
diagrams:
\begin{equation}
\begC1{\phi_{xy}^t}\endC1{\phi_{ij}^a}\doteq\delta_{xj}\delta_{iy}\,\delta^{ta}
\end{equation}
The remaining fields contract together to give the propagator of
Eq.~(\ref{prop}) in the loop, and integrating over the loop momentum gives the
integral of Eq.~(\ref{Kintdef}):
\begin{align}
\begC1{\phi_{ij}^a}\endC1{\phi_{kl}^b}&=
\delta^{ab}\left(\delta_{il}\delta_{jk}\frac{1}{q^2+\frac{1}{2}(I_a+J_a)}+\delta_{ij}\delta_{kl}D^a_{il}\right)\nonumber\\
&\rightarrow \delta^{ab}\,K_{ij,kl}^{a}\quad\text{no sum}\nonumber
\end{align}
There are six ways to contract the vertex fields and external fields while
maintaining the order of the latter, or equivalently, while maintaining the
order of $x$ and $y$.  Suppressing the common factors of the coupling
$\mu/(48f^2)$, taste factor $\tau_{abcd}$, and mass $m_i$, we have
\begin{align}
\delta_{xj,iy,yk,jx}^{ta,tb,cd}\,K_{kl,li}^{c}
&+\delta_{xj,iy,yl,kx}^{ta,tc,bd}\,K_{jk,li}^{b}\nonumber\\
+\,\delta_{xj,iy,yi,lx}^{ta,td,bc}\,K_{jk,kl}^{b}
&+\delta_{xk,jy,yl,kx}^{tb,tc,ad}\,K_{ij,li}^{a}\nonumber\\
+\,\delta_{xk,jy,yi,lx}^{tb,td,ac}\,K_{ij,kl}^{a}
&+\delta_{xl,ky,yi,lx}^{tc,td,ab}\,K_{ij,jk}^{a}\nonumber\\
+\,(x&\leftrightarrow y)\label{eq:contrct_delt_K}
\end{align}
where $\delta$ is simply a product of Kronecker deltas:
\begin{equation}
\delta_{ij,kl,mn,pq}^{ab,cd,ef}\equiv\delta_{ij}\delta_{kl}\delta_{mn}\delta_{pq}\,\delta^{ab}\delta^{cd}\delta^{ef}.
\end{equation}
Restoring the taste factor $\tau_{abcd}$ and quark mass $m_i$ and summing over
the flavor and taste indices gives
\begin{align}
\tau_{ttcc}\,m_y\,K_{yi,iy}^{c}
&+\tau_{tbtb}\,m_y\,K_{xx,yy}^{b}\nonumber\\
+\,\tau_{tbbt}\,m_y\,K_{xi,ix}^{b}
&+\tau_{atta}\,m_i\,K_{yi,iy}^{a}\nonumber\\
+\,\tau_{atat}\,m_y\,K_{xx,yy}^{a}
&+\tau_{aatt}\,m_y\,K_{yi,iy}^{a}\nonumber\\
+\,(x&\leftrightarrow y)\label{eq:tau_m_K}
\end{align}
where we used the symmetry of the propagator under interchange of the fields
and relabeled dummy flavor indices.  The taste matrices $T^a$ all commute or
anticommute with one another.  Defining $\theta^{ab}$ such that
\begin{equation}
T^aT^b=\theta^{ab}\,T^bT^a\quad\forall\quad a,b\label{eq:def_theta}
\end{equation}
and noting
\begin{align}
(T^a)^2&=\xi_I\quad\forall\quad a,\label{eq:sqr_T}
\end{align}
Eq.~(\ref{eq:tau_m_K}) becomes
\begin{align}
4\sum_a\biggl[(2m_x+m_y+m_i)K_{xi,ix}^a\nonumber\\
+(2m_y+m_x+m_i)K_{yi,iy}^a\nonumber\\
+2\theta^{at}(m_x+m_y)K_{xx,yy}^a\biggr].
\end{align}
Restoring the coupling gives the desired result, Eq.~(\ref{mass4ptloops}).  

Contributions from loops with kinetic energy, mass, and hairpin vertices
(respectively Eqs.~(\ref{KE444}), (\ref{mass444}), and (\ref{hair444})) cancel
against the term with $\theta^{at}$ in Eq.~(\ref{mass4ptloops}), and
evaluations of the sums over $\theta^{at}$ (within irreps of $SO(4)_T$) are not
needed to arrive at the coefficient of the mass term in the result,
Eq.~(\ref{disc444}).  In general such cancellations do not occur.  The needed
sums are collected in Appendix~\ref{sum_theta}.

\subsection{Potential $a^2\mathcal{U}$ vertices}
We begin by noting that the flavor structure of the vertices from
$a^2\mathcal{U}$ is identical to the flavor structure of the mass vertices
(Eq.~(\ref{m4pt_rem})):
\begin{align}
-\frac{a^2C_1}{12f^4}\,&(\tau_{abcd}+3\tau_{5ab5cd}-4\tau_{5a5bcd})\,\phi^a_{ij}\phi^b_{jk}\phi^c_{kl}\phi^d_{li}\label{C1_4pt_rem}\\
-\frac{a^2C_6}{12f^4}\,&(\tau_{abcd}+3\tau_{\mu\nu,ab,\mu\nu,cd}-4\tau_{\mu\nu,a,\mu\nu,bcd})\,\phi^a_{ij}\phi^b_{jk}\phi^c_{kl}\phi^d_{li} \label{C6_4pt_rem}\\
-\frac{a^2C_3}{12f^4}\,&(\tau_{abcd}+3\tau_{\nu ab\nu cd}+4\tau_{\nu a\nu bcd})\,\phi^a_{ij}\phi^b_{jk}\phi^c_{kl}\phi^d_{li} \label{C3_4pt_rem}\\
-\frac{a^2C_4}{12f^4}\,&(\tau_{abcd}+3\tau_{\nu5,ab,\nu5,cd}+4\tau_{\nu5,a,\nu5,bcd})\,\phi^a_{ij}\phi^b_{jk}\phi^c_{kl}\phi^d_{li} \label{C4_4pt_rem}.
\end{align}
Only the overall normalizations (couplings) and taste factors differ from those
in Eq.~(\ref{m4pt_rem}).  Therefore the same contractions and corresponding
products of Kronecker deltas and integrals that appear in
Eq.~(\ref{eq:contrct_delt_K}) enter the calculation, and we can obtain the
loops for each of the vertex classes in Eqs.~(\ref{C1_4pt_rem}) through
(\ref{C4_4pt_rem}) from Eqs.~(\ref{eq:contrct_delt_K}) and (\ref{eq:tau_m_K})
by taking $m_i\rightarrow 1$ and replacing $\tau_{abcd}$ in
Eq.~(\ref{eq:tau_m_K}) with the appropriate linear combinations of traces from
Eqs.~(\ref{C1_4pt_rem}) through (\ref{C4_4pt_rem}).

Noting that the taste factors in Eqs.~(\ref{C1_4pt_rem}) through (\ref{C4_4pt_rem}) all have the same form, {\it viz.}
\begin{equation}
\tau_{abcd}+3\tau_{sabscd}-4\theta^{5s}\tau_{sasbcd}
\end{equation}
where
\begin{align}
s=
\begin{cases}
5 & \text{for vertices $\propto C_1$}\\
\mu\nu & \text{for vertices $\propto C_6$}\\
\nu & \text{for vertices $\propto C_3$}\\
\nu5 & \text{for vertices $\propto C_4$},
\end{cases}
\end{align}
and recalling Eqs.~(\ref{eq:def_theta}) and (\ref{eq:sqr_T}), we find the taste factors in Eqs.~(\ref{C1_4pt_rem}) through (\ref{C4_4pt_rem}) are
\begin{align}
\tau_{ttcc}+3&\tau_{sttscc}-4\theta^{5s}\tau_{ststcc}=16(1-\theta^{5s}\theta^{st})\nonumber\\
\tau_{tbtb}+3&\tau_{stbstb}-4\theta^{5s}\tau_{stsbtb}\nonumber\\
&=4\theta^{bt}(1+3\theta^{bs}\theta^{st}-4\theta^{5s}\theta^{st})\nonumber\\
\tau_{tbbt}+3&\tau_{stbsbt}-4\theta^{5s}\tau_{stsbbt}=4(1+3\theta^{bs}\theta^{st}-4\theta^{5s}\theta^{st})\nonumber\\
\tau_{atta}+3&\tau_{satsta}-4\theta^{5s}\tau_{sastta}=4(1+3\theta^{as}\theta^{st}-4\theta^{5s}\theta^{sa})\nonumber\\
\tau_{atat}+3&\tau_{satsat}-4\theta^{5s}\tau_{sastat}\nonumber\\
&=4\theta^{at}(1+3\theta^{as}\theta^{st}-4\theta^{5s}\theta^{sa})\nonumber\\
\tau_{aatt}+3&\tau_{saastt}-4\theta^{5s}\tau_{sasatt}=16(1-\theta^{5s}\theta^{sa}).\nonumber
\end{align}
Replacing the taste factors in Eq.~(\ref{eq:tau_m_K}) with these (and taking
$m_i\rightarrow 1$ there) leads directly to Eq.~(\ref{C4ptloops}).
%
%
%
%

\section{\label{sum_theta}Coefficient calculations}
Here we detail one way to calculate the coefficients in Tables~\ref{ConDiag2},
\ref{ConDiag}, and \ref{DiscDiag}.  The coefficients defined in
Eqs.~\eqref{dat_def} and \eqref{eat_def} can be computed similarly.  As a
by-product of our calculation, we obtain the intermediate results of
Tables~\ref{table:consums} and~\ref{table:discsums}, which we calculated twice:
Once using the (anti)commutation relations of the generators to count the
positive and negative terms in each sum, and once using the relation
$\theta^{ab}=\tau_{abab}/4$ and a computer.  We have also checked
Table~\ref{table:consums} (\ref{table:discsums}) implicitly, using an
independent accounting scheme to arrive at the coefficients $\Delta_{BF}^{con}$
($\Delta_{BF}^{disc}$) given in Table~\ref{ConDiag} (\ref{DiscDiag}).

The coefficients $\delta_{BF}^{con}$ are defined in Eq.~(\ref{delBFcon_def}):
\begin{align}
\delta_{BF}^{con}&=\frac{1}{32}\sum_{a\in B}\sum_{b\in V,A}\omega_b\tau_{abt}\tau_{abt}(1+\theta^{ab})\\
&=\frac{1}{32}\sum_{\mu,a\in B}\bigl[\omega_V(\tau_{a\mu t})^2(1+\theta^{a\mu})\\
&\phantom{=\frac{1}{32}\sum_{\mu,a\in B}}+ \omega_A(\tau_{a,\mu5,t})^2(1+\theta^{a,\mu5})\bigr],
\end{align}
and we note
\begin{align}
1+\theta^{a,\mu(5)}=
\begin{cases}
2 & \text{if $[T^a,\xi_{\mu(5)}]=0$}\\
0 & \text{if $\{T^a,\xi_{\mu(5)}\}=0$,}
\end{cases}
\end{align}
so that 
\begin{equation}
\delta_{PF}^{con}=0\quad\forall\ F
\end{equation}
and
\begin{align}
\delta_{IF}^{con}
&=\frac{1}{16}\sum_\mu\bigl[\omega_V(\tau_{\mu t})^2 + \omega_A(\tau_{\mu5,t})^2\bigr]\\
&=\sum_\mu\bigl[\omega_V\delta^{\mu t} + \omega_A\delta^{\mu5,t}\bigr]\\
\delta_{IF}^{con}&=
\begin{cases}
\omega_V & \text{if $F=V$}\\
\omega_A & \text{if $F=A$}\\
0 & \text{otherwise.}
\end{cases}
\end{align}
%
%
Similarly, for $B=V$ we have
\begin{align}
\delta_{VF}^{con}
&=\frac{1}{32}\sum_{\mu,\nu}\bigl[\omega_V(\tau_{\nu\mu t})^2(1+\theta^{\nu\mu})\\
&\phantom{=\frac{1}{32}\sum_{\mu,\nu}}+ \omega_A(\tau_{\nu,\mu5,t})^2(1+\theta^{\nu,\mu5})\bigr]\\
&=\frac{1}{16}\sum_{\mu}\,\omega_V(\tau_{t})^2
+\frac{1}{8}\sum_{\mu<\nu}\,\omega_A(\tau_{\nu,\mu5,t})^2\\
\delta_{VF}^{con}&=
\begin{cases}
4\omega_V & \text{if $F=I$}\\
2\omega_A & \text{if $F=T$}\\
0 & \text{otherwise,}
\end{cases}
\end{align}
and for $B=A$, we have
\begin{align}
\delta_{AF}^{con}
&=\frac{1}{32}\sum_{\mu,\nu}\bigl[\omega_V(\tau_{\nu5,\mu t})^2(1+\theta^{\nu5,\mu})\\
&\phantom{=\frac{1}{32}\sum_{\mu,\nu}}+ \omega_A(\tau_{\nu5,\mu5,t})^2(1+\theta^{\nu5,\mu5})\bigr]\\
&=\frac{1}{32}\sum_{\mu,\nu}\bigl[\omega_V(\tau_{\nu,\mu5, t})^2(1+\theta^{\mu5,\nu})\\
&\phantom{=\frac{1}{32}\sum_{\mu,\nu}}+ \omega_A(\tau_{\nu\mu t})^2(1+\theta^{\nu\mu})\bigr]\\
\delta_{AF}^{con}&=
\begin{cases}
2\omega_V & \text{if $F=T$}\\
4\omega_A & \text{if $F=I$}\\
0 & \text{otherwise.}
\end{cases}
\end{align}
Finally, for $B=T$ we have
\begin{align}
\delta_{TF}^{con}
&=\frac{1}{32}\sum_{\mu,\rho<\lambda}\bigl[\omega_V(\tau_{\rho\lambda,\mu t})^2(1+\theta^{\rho\lambda,\mu})\\
&\phantom{=\frac{1}{32}\sum_{\mu,a\in B}}+ \omega_A(\tau_{\rho\lambda,\mu5,t})^2(1+\theta^{\rho\lambda,\mu5})\bigr]\\
&=\frac{1}{16}\sum_{\stackrel{\rho<\lambda}{\mu\neq\rho,\lambda}}\bigl[\omega_V(\tau_{\rho\lambda,\mu t})^2
+ \omega_A(\tau_{\rho\lambda,\mu5,t})^2\bigr]\label{eq:commute}\\
\delta_{TF}^{con}&=
\begin{cases}
3\omega_V & \text{if $F=A$}\\
3\omega_A & \text{if $F=V$}\\
0 & \text{otherwise.}
\end{cases}\label{eq:delTF}
\end{align}
These results are straightforwardly obtained by substituting for $t$ and
counting the number of nonzero terms in the sums.  For evaluating the traces,
the (anti)commutation of the generators $T^a$, the fact that $(T^a)^2=\xi_I$, the
orthogonality relations $\Tr(T^aT^b)=4\delta^{ab}$, and the traces over
products of Euclidean gamma matrices $\gamma_\mu$ are useful.
%
%

\begin{table}[tbhp]
\begin{center}
\caption{
  \label{table:consums}Sums for evaluating the coefficients
  $\Delta_{BF}^{con}$.  
  All the sums required for the
  coefficients $\Delta_{BF}^{con}$ can be obtained by repeated use of
  the results in the first three lines.  }
\begin{tabular}{r|rrrrr}
\hline\hline
$t\in$ & $P$ & $I$ & $V$ & $A$ & $T$ \\
\hline
$\sum_\mu\,\theta^{\mu t}$ & $-4$ & $4$ & $-2$ & $2$ & $0$ \\
$\sum_\mu\,\theta^{\mu5,t}$ & $-4$ & $4$ & $2$ & $-2$ & $0$ \\
$\sum_{\mu<\nu}\,\theta^{\mu\nu,t}$ & $6$ & $6$ & $0$ & $0$ & $-2$ \\
$\sum_{\rho,\mu<\nu}\,\theta^{\mu\nu,\rho}\theta^{\mu\nu,t}$ & $0$ & $0$ & $0$ & $0$ & $0$ \\
$\sum_{\rho,\mu}\,\theta^{\mu\rho}\theta^{\mu t}$ & $8$ & $-8$ & $4$ & $-4$ & $0$ \\
$\sum_{\rho,\mu}\,\theta^{\mu\rho}\theta^{\mu5,t}$ & $8$ & $-8$ & $-4$ & $4$ & $0$ \\
$\sum_{\rho<\lambda,\mu<\nu}\,\theta^{\mu\nu,\rho\lambda}\theta^{\mu\nu,t}$ & $-12$ & $-12$ & $0$ & $0$ & $4$ \\
$\sum_{\rho<\lambda,\mu}\,\theta^{\mu,\rho\lambda}\theta^{\mu t}$ & $0$ & $0$ & $0$ & $0$ & $0$ \\
$\sum_{\rho<\lambda,\mu}\,\theta^{\mu,\rho\lambda}\theta^{\mu5,t}$ & $0$ & $0$ & $0$ & $0$ & $0$ \\
\hline\hline
\end{tabular}
\end{center}
\end{table}

The coefficients of Table~\ref{ConDiag} are defined in
Eq.~(\ref{DelBFcon_def}):
\begin{equation}
\Delta_{BF}^{con}=\sum_{a\in B}\biggl(\Delta_{at}-(\Delta_t+\Delta_a)\biggr).
\end{equation}
Substituting for the coefficients $\Delta_{at}$ and taste splittings using Eqs.~(\ref{dat_def}) and (\ref{reduce}) gives
\begin{align}
\Delta_{BF}^{con}=
\frac{24}{f^2}\sum_{\stackrel{a\in B}{b\neq I}}&C_b(1+\theta^{ab}\theta^{bt}-\theta^{5b}\theta^{bt}-\theta^{ab}\theta^{b5})\label{eq:DelBFcon}\\
=
\frac{24}{f^2}\sum_{a\in B}\biggl(C_1&(1+\theta^{5a}\theta^{5t}-\theta^{5t}-\theta^{5a})\\
+\sum_{\mu<\nu}C_6&(1+\theta^{\mu\nu,a}\theta^{\mu\nu,t}-\theta^{\mu\nu,t}-\theta^{\mu\nu,a})\\
+\sum_\mu C_3&(1+\theta^{\mu a}\theta^{\mu t}+\theta^{\mu t}+\theta^{\mu a})\\
+\sum_\mu C_4&(1+\theta^{\mu5,a}\theta^{\mu5,t}+\theta^{\mu5,t}+\theta^{\mu5,a})\biggr).
\end{align}
Writing out the sum over each irrep $B$ gives
\begin{align}
\Delta_{PF}^{con}&=0\label{eq:DeltaPFcon}\\
\Delta_{IF}^{con}&=
\frac{48}{f^2}\sum_\mu\bigl(C_3(1+\theta^{\mu t})+
C_4(1+\theta^{\mu5,t})\bigr)\label{eq:DeltaIFcon}\\
\Delta_{VF}^{con}&=
\frac{24}{f^2}\sum_{\rho}\biggl(2C_1(1-\theta^{5t})\\
&+\sum_{\mu<\nu}C_6(1+\theta^{\mu\nu,\rho}\theta^{\mu\nu,t}-\theta^{\mu\nu,t}-\theta^{\mu\nu,\rho})\\
&+\sum_\mu C_3(1+\theta^{\mu \rho}\theta^{\mu t}+\theta^{\mu t}+\theta^{\mu \rho})\\
&+\sum_\mu C_4(1-\theta^{\mu\rho}\theta^{\mu5,t}+\theta^{\mu5,t}-\theta^{\mu\rho})\biggr)\\
\Delta_{AF}^{con}&=
\frac{24}{f^2}\sum_{\rho}\biggl(2C_1(1-\theta^{5t})\\
&+\sum_{\mu<\nu}C_6(1+\theta^{\mu\nu,\rho}\theta^{\mu\nu,t}-\theta^{\mu\nu,t}-\theta^{\mu\nu,\rho})\\
&+\sum_\mu C_3(1-\theta^{\mu\rho}\theta^{\mu t}+\theta^{\mu t}-\theta^{\mu\rho})\\
&+\sum_\mu C_4(1+\theta^{\mu\rho}\theta^{\mu5,t}+\theta^{\mu5,t}+\theta^{\mu\rho})\biggr)\\
\Delta_{TF}^{con}&=
\frac{24}{f^2}\sum_{\rho<\lambda}\Biggl(
\sum_{\mu<\nu}C_6\times\\
&\phantom{=\frac{24}{f^2}}(1+\theta^{\mu\nu,\rho\lambda}\theta^{\mu\nu,t}-\theta^{\mu\nu,t}-\theta^{\mu\nu,\rho\lambda})\\
&+\sum_\mu C_3(1+\theta^{\mu,\rho\lambda}\theta^{\mu t}+\theta^{\mu t}+\theta^{\mu,\rho\lambda})\\
&+\sum_\mu C_4(1+\theta^{\mu,\rho\lambda}\theta^{\mu5,t}+\theta^{\mu5,t}+\theta^{\mu,\rho\lambda})\Biggr).\label{eq:DeltaTFcon}
\end{align}
Inspecting Eqs.~(\ref{eq:DeltaIFcon}) through (\ref{eq:DeltaTFcon}), we note
the requisite sums.  Their values are given in Table~\ref{table:consums}.
Using Table~\ref{table:consums} to evaluate Eqs.~(\ref{eq:DeltaIFcon}) through
(\ref{eq:DeltaTFcon}) yields the results of Table~\ref{ConDiag}.  For $F=P\
(t=5)$, all the coefficients explicitly vanish, as they must.
%
%

\begin{table}[tbhp]
\begin{center}
  \caption{ \label{table:discsums}Sums (in addition to those in
    Table~\ref{table:consums}) for evaluating the coefficients
    $\Delta_{BF}^{disc}$.  
    The last three lines can be obtained from the first two, since
    $\theta^{\mu5,t}=\theta^{5t}\theta^{\mu t}$.  }
\begin{tabular}{r|rrrrr}
\hline\hline
$t\in$ & $P$ & $I$ & $V$ & $A$ & $T$ \\
\hline
$\sum_{\rho,\mu<\nu}\,\theta^{\rho t}\theta^{\mu\nu,\rho}\theta^{\mu\nu,t}$ & $0$ & $0$ & $12$ & $-12$ & $0$ \\
$\sum_{\rho,\mu}\,\theta^{\rho t}\theta^{\mu\rho}\theta^{\mu t}$ & $-8$ & $-8$ & $4$ & $4$ & $8$ \\
$\sum_{\rho<\lambda,\mu<\nu}\,\theta^{\rho\lambda,t}\theta^{\mu\nu,\rho\lambda}\theta^{\mu\nu,t}$ & $-12$ & $-12$ & $0$ & $0$ & $20$ \\
$\sum_{\rho,\mu}\,\theta^{\rho t}\theta^{\mu\rho}\theta^{\mu5,t}$ & $-8$ & $-8$ & $-4$ & $-4$ & $8$ \\
$\sum_{\rho,\mu<\nu}\,\theta^{\rho5,t}\theta^{\mu\nu,\rho}\theta^{\mu\nu,t}$ & $0$ & $0$ & $-12$ & $12$ & $0$ \\
$\sum_{\rho,\mu}\,\theta^{\rho5,t}\theta^{\mu\rho}\theta^{\mu5,t}$ & $-8$ & $-8$ & $4$ & $4$ & $8$ \\
\hline\hline
\end{tabular}
\end{center}
\end{table}

The coefficients of Table~\ref{DiscDiag} are defined in Eq.~(\ref{DeltaDBFDef}):
\begin{equation}
\Delta_{BF}^{disc}=\sum_{a\in B}\biggl(\Delta^\prime_{at}+\theta^{at}\Delta_t + (1+\rho^{at}/2)\Delta_a\biggr).
\end{equation}
Substituting for the coefficients $\Delta^\prime_{at}$, taste splittings, and coefficients $\rho^{at}$ using Eqs.~\eqref{eat_def}, \eqref{reduce} and \eqref{dat_def}, and \eqref{eq:rho} gives, for $B\ne I$,
\begin{equation}
\Delta_{BF}^{disc}=
\frac{24}{f^2}\sum_{\stackrel{a\in B}{b\neq I}}C_b\bigl(-1+\theta^{at}(\theta^{ab}\theta^{bt}-\theta^{5b}\theta^{bt})+\theta^{ab}\theta^{b5}\bigr),\label{eq:DelBFdisc_BnotI}
\end{equation}
and for $B=I$,
\begin{equation}
\Delta_{IF}^{disc}=
\frac{24}{f^2}\sum_{b\ne I}C_b\bigl(1+\theta^{bt}-\theta^{5b}\theta^{bt}-\theta^{b5}\bigr).\label{eq:DelBFdisc_BisI}
\end{equation}
For $B\ne I$, adding and subtracting Eqs.~\eqref{eq:DelBFdisc_BnotI} and \eqref{eq:DelBFcon} gives
\begin{align}
\Delta_{BI}^{disc}&=\Delta_{BI}^{con}-6N_B\Delta_B\quad(B\ne I)\label{eq:DelBIdisc_chk}\\
\Delta_{PT}^{disc}&=\Delta_{PT}^{con}\\
\Delta_{TP}^{disc}&=\Delta_{TP}^{con}-36\Delta_T\\
\Delta_{PV}^{disc}&=-\Delta_{PV}^{con}\\
\Delta_{PA}^{disc}&=-\Delta_{PA}^{con}\\
\Delta_{VP}^{disc}&=-\Delta_{VP}^{con}\label{eq:DelVPdisc_chk}\\
\Delta_{AP}^{disc}&=-\Delta_{AP}^{con},\label{eq:DelAPdisc_chk}
\end{align}
while for $B=I$, comparing Eqs.~\eqref{eq:DelBFdisc_BisI} and \eqref{eq:DelBFcon} gives
\begin{align}
\Delta_{IF}^{disc}=\Delta_{IF}^{con}.\label{eq:DelIFdisc}
\end{align}
Eq.~\eqref{eq:DelBIdisc_chk} implies Eqs.~\eqref{eq:DelVIdisc_chk} and
\eqref{eq:DelAIdisc_chk}; they and Eqs.~\eqref{eq:DelVPdisc_chk} and
\eqref{eq:DelAPdisc_chk} can be used to cross-check the results in
Table~\ref{DiscDiag}.  Eq.~\eqref{eq:DelIFdisc} and Table~\ref{ConDiag} give
the coefficients $\Delta_{IF}^{disc}$, while the coefficients
$\Delta_{PF}^{disc}$ and $\Delta_{TF}^{disc}$ do not appear in Eq.~\eqref{disc444};
the remaining coefficients in Eq.~\eqref{disc444} are $\Delta_{VF}^{disc}$
and $\Delta_{AF}^{disc}$.

Writing out the sum over $b$ in Eq.~\eqref{eq:DelBFdisc_BnotI} gives
\begin{align}
&\Delta_{BF}^{disc}=
\frac{24}{f^2}\sum_{a\in B}\biggl(C_1\bigl(-1+\theta^{at}(\theta^{5a}\theta^{5t}-\theta^{5t})+\theta^{5a}\bigr)\label{eq:DelBFdisc_C1}\\
&+\sum_{\mu<\nu}C_6\bigl(-1+\theta^{at}(\theta^{\mu\nu,a}\theta^{\mu\nu,t}-\theta^{\mu\nu,t})+\theta^{\mu\nu,a}\bigr)\\
&+\sum_\mu C_3\bigl(-1+\theta^{at}(\theta^{\mu a}\theta^{\mu t}+\theta^{\mu t})-\theta^{\mu a}\bigr)\\
&+\sum_\mu C_4\bigl(-1+\theta^{at}(\theta^{\mu5,a}\theta^{\mu5,t}+\theta^{\mu5,t})-\theta^{\mu5,a}\bigr)\biggr),\label{eq:DelBFdisc_C4}
\end{align}
and writing out the sums over the vector and axial irreps ($B=V$ and $B=A$) gives
\begin{align}
\Delta&_{VF}^{disc}=
\frac{24}{f^2}\sum_{\rho}\biggl(2C_1(-1-\theta^{\rho t}\theta^{5t})\label{eq:DelVFdisc}\\
&+\sum_{\mu<\nu}C_6\bigl(-1+\theta^{\rho t}(\theta^{\mu\nu,\rho}\theta^{\mu\nu,t}-\theta^{\mu\nu,t})+\theta^{\mu\nu,\rho}\bigr)\\
&+\sum_\mu C_3\bigl(-1+\theta^{\rho t}(\theta^{\mu \rho}\theta^{\mu t}+\theta^{\mu t})-\theta^{\mu \rho}\bigr)\\
&+\sum_\mu C_4\bigl(-1-\theta^{\rho t}(\theta^{\mu\rho}\theta^{\mu5,t}-\theta^{\mu5,t})+\theta^{\mu\rho}\bigr)\biggr),
\end{align}
\begin{align}
\Delta&_{AF}^{disc}=
\frac{24}{f^2}\sum_{\rho}\biggl(2C_1(-1-\theta^{\rho5,t}\theta^{5t})\\
&+\sum_{\mu<\nu}C_6\bigl(-1+\theta^{\rho5,t}(\theta^{\mu\nu,\rho}\theta^{\mu\nu,t}-\theta^{\mu\nu,t})+\theta^{\mu\nu,\rho}\bigr)\\
&+\sum_\mu C_3\bigl(-1-\theta^{\rho5,t}(\theta^{\mu\rho}\theta^{\mu t}-\theta^{\mu t})+\theta^{\mu\rho}\bigr)\\
&+\sum_\mu C_4\bigl(-1+\theta^{\rho5,t}(\theta^{\mu\rho}\theta^{\mu5,t}+\theta^{\mu5,t})-\theta^{\mu\rho}\bigr)\biggr).\label{eq:DelAFdisc}
\end{align}
Examining Eqs.~\eqref{eq:DelVFdisc} through \eqref{eq:DelAFdisc}, we note the
sums beyond those in Table~\ref{table:consums} that are needed to evaluate the
coefficients $\Delta_{VF}^{disc}$ and $\Delta_{AF}^{disc}$.  The values of
these sums are given in Table~\ref{table:discsums}.  Using the sums in
Tables~\ref{table:consums} and \ref{table:discsums} in
Eqs.~\eqref{eq:DelVFdisc} through \eqref{eq:DelAFdisc} yields the results in
Table~\ref{DiscDiag}.  From Eq.~\eqref{eq:DelBFdisc_BisI} and
Eqs.~\eqref{eq:DelBFdisc_C1} through \eqref{eq:DelBFdisc_C4}, we see that $\Delta_{IP}^{disc}=\Delta_{VP}^{disc}=\Delta_{AP}^{disc}=0$,
as necessary for the result in Eq.~\eqref{disc444} to reduce properly in the taste
Goldstone case.
%
%
%
%
%
\bibliographystyle{apsrev} 
\bibliography{ref} 
\end{document}